\documentclass[aps,twocolumn,showpacs,preprintnumbers]{revtex4}
\usepackage{graphics}

\newcommand{\beq}{\begin{equation}}
\newcommand{\eeq}{\end{equation}}
\newcommand{\ber}{\begin{eqnarray}}
\newcommand{\eer}{\end{eqnarray}}
\newcommand{\berr}{\begin{eqnarray*}}
\newcommand{\eerr}{\end{eqnarray*}}
\newcommand{\bt}{\beta}
\newcommand{\kp}{\kappa}
\newcommand{\kc}{\kappa_c}

\newcommand{\sg}{\sigma(\omega,\vec{p})}
\newcommand{\sgh}{\sigma_H(p_0,\vec{p})}
\newcommand{\sw}{\sigma(\omega)}
\newcommand{\rhw}{\rho(\omega)}

\newcommand{\om}{\omega}
\newcommand{\tc}{T_c}
\newcommand{\ec}{\eta_c}
\newcommand{\jpsi}{J / \psi}
\newcommand{\as}{\chi_c^0}
\newcommand{\ax}{\chi_c^1}
\newcommand{\al}{\alpha}
\newcommand{\mpl}{m_{\rm pole}}
\newcommand{\mkin}{m_{\rm kinetic}}

\newcommand{\qq} {q \bar{q}}
\newcommand{\grecon} {G_{\rm recon}(\tau, T)}
\newcommand{\gratio} {G(\tau, T) / G_{\rm recon}(\tau, T)}
\newcommand{\simgt}{\,\rlap{\lower 3.5 pt\hbox{$\mathchar \sim$}}\raise 1pt \hbox{$>$}\,}
\newcommand{\simlt}{\,\rlap{\lower 3.5 pt\hbox{$\mathchar \sim$}}\raise 1pt \hbox{$<$}\,}

\begin{document}
\title{Behavior of Charmonium systems after Deconfinement}
\author{Saumen Datta and Frithjof Karsch}
\email{saumen,karsch@physik.uni-bielefeld.de}
\affiliation{Fakult\"at f\"ur Physik, Universit\"at Bielefeld, 
D33615 Bielefeld, Germany.}
\author{Peter Petreczky}
\email{petreczk@quark.phy.bnl.gov}
\thanks{Goldhaber and RIKEN Fellow}
\affiliation{Physics Department, Brookhaven National Laboratory,
Upton, NY 11973, USA}
\author{Ines Wetzorke}
\email{Ines.Wetzorke@desy.de}
\affiliation{NIC/DESY Zeuthen, Platanenallee 6, D-15738 Zeuthen, Germany.}

\begin{abstract}
We present a study of charmonia in hot gluonic plasma, for
temperatures upto three times the deconfinement transition
temperature $\tc$. $\qq$ systems with quark masses
close to the charm mass and different spin-parity quantum
numbers were studied on very fine isotropic lattices. 
The analysis of temporal correlators, and spectral functions
constructed from them, shows that the $\jpsi$ and $\ec$
survive up to quite high temperatures, with little
observable change up to 1.5 $\tc$, and then gradually
weakening and disappearing by 3 $\tc$. 
For the scalar and axial vector channels, serious
modifications are induced by the hot medium already close
to $\tc$, possibly dissociating the mesons by 1.1 $\tc$.
\end{abstract}
\pacs{11.15.Ha, 12.38.Gc, 12.38.Mh, 25.75.Nq}
\preprint{BI-TP 2003/39, DESY 03-215, hep-lat/0312037}
\maketitle

\section{INTRODUCTION}
\label{sec.introduction}

The behavior of strongly interacting matter in a hot and
dense environment has been a subject of considerable
theoretical and experimental research. At very high
temperatures and densities, hadronic matter is expected to undergo a
phase transition (or crossover) to a deconfined plasma
state. Dedicated heavy ion collision experiments are
aiming to create this state, the study of which will enrich our
understanding of strong interactions and which is also directly
relevant for the early universe. The
supression of $\jpsi$ peak in the dilepton spectrum is 
one of the most important signals of this phase transition.  
Unlike light quarks, charmonia may survive as bound states
even in a deconfined medium, due to Coulomb attraction
between the quarks. However,
based on non-relativistic arguments, Matsui and 
Satz predicted that already at temperatures close to 
$\tc$, binding between quarks is reduced enough to dissolve
$J / \psi$, and proposed its suppression as a signal of the 
deconfinement transition \cite{satz}. 
Several later studies, based on potential model calculations,
predicted a pattern of dissolution, with the higher
excitations dissolving earlier, and $J / \psi$ dissolving
at a temperature $\approx$ 1.1 $\tc$ \cite{karsch,digal}. 

Since the validity of such potential model calculations 
at high temperatures are not a priori clear, a more reliable way of
studying properties of charmonia in a hot medium is clearly 
desirable. One possibility is to directly analyze the
thermal Green functions of the corresponding states.
For a long time it has been thought that numerical study
on the lattice is not a practical tool for calculations 
of dynamic properties of QCD at finite temperature. 
However, it was shown recently that meson spectral
functions, which are directly related to the real time
correlators, can be extracted from temporal
correlators calculated on the lattice using the Maximum
Entropy Method (MEM) \cite{nakahara99,asakawa01}.
Some earlier investigations of meson spectral functions
(without the use of MEM) were presented in \cite{qcdtaro}. 
The method has been successfully applied at zero temperature
\cite{nakahara99,asakawa01,wetzorke00,yamazaki01}.
Additional difficulties are present at finite temperature
because of the finite physical extent of the Euclidean time
direction. However, the methods of Ref. \cite{asakawa01}
were applied successfully to gain qualitative information
about the medium modification of mesonic states with
temperature \cite{wetzorke01,karsch01,
asakawa02}. Such studies have provided useful 
quantitative information about dilepton 
rates \cite{karsch01}, and quite unexpected nontrivial
structure of the low energy spectral function for the
$\bar{s} s$ mesonic states \cite{asakawa02}.  

First applications of such methods to charmonium systems 
also produced interesting and unexpected results
\cite{lat02,umeda02}. The ground state charmonia, 
$\ec$ and $\jpsi$, were found to survive 
well after the deconfinement transition, at least upto
temperatures of 1.5 $\tc$ \cite{lat02}, and no significant
modifications of their masses were found on crossing the
transition temperature \cite{umeda02}. Both these features
are in sharp contrast to the existing potential model
studies. The 1P states $\as$ and $\ax$, on the other 
hand, were found to undergo serious temperature 
modifications, possibly dissolution, already very 
close to $\tc$ \cite{lat02}. Later studies confirmed 
these results, and also found evidence of disappearance 
of the 1S states from the spectral function at 
higher temperatures \cite{asakawa03,lat03}.

In the present paper we are going to report a detailed
study of finite temperature charmonium correlators and spectral 
functions in quenched QCD using very fine isotropic
lattices, expanding on preliminary results published 
in \cite{lat02,lat03}. The plan of the paper is as
follows. In the next section, we give a short discussion of
the definition of the spectral function and its properties.
Sec. \ref{sec.lattices} gives details of
our lattices and simulation parameters. In Sec. \ref{sec.belowtc} we
explain our analysis methods, giving a short outline of
the maximum entropy method. We then present the results
below the deconfining transition, which allows us to
explain some features of our analysis that are used
later. Sec. \ref{sec.abovetc} is the central part of
the paper, where we present our results for the charmonium
correlators and spectral functions above $\tc$. Study of
the systematics is a crucial part of the MEM analysis, at least at the
current stage of the finite temperature mesonic spectral
function studies. Therefore, a detailed discussion of the 
dependence of the results of Sec. \ref{sec.abovetc} on 
different systematics follows in Sec. \ref{sec.systematics}. In the
next section, we discuss the spatial correlators and
screening masses of charmonia, which show some interesting
thermal effects and lead us to in-medium modifications of
the dispersion relations for the $\jpsi$. Finally, 
Sec. \ref{sec.summary} contains a summary of our results, 
along with their phenomenological implications. We also
shortly discuss here some points regarding the potential
model studies. Readers who are interested in
heavy ion phenomenology but not particularly interested in
the details of our analysis may concentrate on Sections
\ref{sec.abovetc} and \ref{sec.summary}.

\section{SPECTRAL FUNCTION AND STATES AT FINITE TEMPERATURE}
\label{sec.states}

Most dynamic properties of thermal systems are incorporated 
in the spectral function. The spectral function $\sgh$ for a given 
mesonic channel $H$ in a system at temperature T can be defined 
through the Fourier transform of the real time two point functions
$D^{>}$ and $D^{<}$, or equivalently, as the imaginary part of 
the Fourier transformed retarded correlation function \cite{lebellac},
\ber
\sgh &=& \frac{1}{2 \pi} (D^{>}_H(p_0, \vec{p})-D^{>}(p_0, \vec{p})) 
\label{eq.defspect} \\
 D^{>(<)}_H(p_0, \vec{p}) &=& \int{d^4 x \over (2
\pi)^4} e^{i p.x} D^{>(<)}_H(x_0,\vec{x}) \nonumber \\
D^{>}_H(x_0,\vec{x}) &=& \langle
J_H(x_0, \vec{x}) J_H(0, \vec{0}) \rangle \label{eq.def2pt} \\
D^{<}_H(x_0,\vec{x}) &=& 
\langle J_H(0, \vec{0}) J_H(x_0,\vec{x}) \rangle \nonumber
\eer
The correlators $D^{>(<)}_H(x_0,\vec{x})$ satisfy the 
well-known Kubo-Martin-Schwinger
(KMS) condition \cite{lebellac}
\beq
D^{>}_H(x_0,\vec{x})=D^{<}(x_0+i/T,\vec{x})
\label{kms}
\eeq
Inserting a complete set of
states in Eq. (\ref{eq.defspect}) and using Eq. (\ref{kms}), 
one gets the expansion
\ber
\sgh = {(2 \pi)^2 \over Z} \sum_{m,n} &{}& (e^{-E_n / T} \pm e^{-E_m / T}) 
|\langle n | J_H(0) | m \rangle|^2. \nonumber \\
&{}& \delta^4(p_\mu - k^n_\mu + k^m_\mu) 
\label{eq.specdef}
\eer
where Z is the partition function, and 
$k^{n(m)}$ refers to the four-momenta of the state $| n (m) \rangle $.

A stable mesonic state contributes a $\delta$ function-like
peak to the spectral function:
\beq
\sgh = | \langle 0 | J_H | H \rangle |^2 \epsilon(p_0)
\delta(p^2 - m_H^2)
\label{eq.stable}
\eeq
where $m_H$ is the mass of the state. For an unstable
particle one gets a smoother peak, with the peak width
being related to the decay width. For sufficiently small
decay width, a Breit-Wigner form is commonly used. 
As one increases the temperature, due to collision broadening the contribution
of the states in the spectral function change, and at sufficiently high
temperatures, the contribution from a state in the spectral function may 
be sufficiently weakened and broadened that it is not very 
meaningful to speak of it as a resonance 
any more. Such change of contributions of
resonance states and eventual `disappearance of resonances'
in the thermal spectral function has been studied
analytically, for example, in the Nambu-Jona-Lasinio 
model in Ref. \cite{hatsuda}. 
Finally, at very high temperatures one would expect the
spectral function to consist only of a smooth continuum
starting at twice the charm quark mass. 
The spectral function as defined in
Eq. (\ref{eq.specdef}) can be directly accessed by high energy
heavy ion experiments. For example, the spectral function for the vector 
current is directly related to the differential thermal cross section 
for the production of dilepton pairs \cite{braaten}:
\beq
{dW \over dp_0 d^3p} |_{\vec{p}=0} = {5 \al^2 \over 27 \pi^2} 
{1 \over p_0^2 (e^{p_0/T}-1)} \sigma_V (p_0, \vec{p}).
\label{eq.dilepton} \eeq
The presence or absence of a bound state in the spectral function
will manifest itself in the peak structure of the differential 
dilepton rate.

In finite temperature lattice calculations, one calculates
Euclidean time propagators, usually
projected to a given spatial momentum:
\beq
G_H(\tau, \vec{p}) = \int d^3x e^{i \vec{p}.\vec{x}} 
\langle T_{\tau} J_H(\tau, \vec{x}) J_H(0,
\vec{0}) \rangle_T
\label{eq.cor}
\eeq
where $\langle ... \rangle_T$ means thermal average,
and $T_\tau$ means ordering in
Euclidean time $\tau$. This quantity is the analytic continuation
of $D^{>}(x_0,\vec{p})$:
\beq
G_H(\tau,\vec{p})=D^{>}(-i\tau,\vec{p}).
\eeq
Using this equation and the KMS condition one can
easily show that $G_H(\tau,\vec{p})$ is related to the 
spectral
function, Eq. (\ref{eq.defspect}), by an integral equation:
\ber
G_H(\tau, \vec{p}) &=& \int_0^{\infty} d \omega
\sigma_H(\om, \vec{p}) K(\om, \tau) \label{eq.spect} \\
K(\omega, \tau) &=& \frac{\cosh(\omega(\tau-1/2
T))}{\sinh(\omega/2 T)}.
\label{eq.kernel}
\eer
Inverting Eq. (\ref{eq.spect}) one can extract
spectral functions and properties of hadrons from
correlators calculated in lattice QCD. In what follows, we
use Eq. (\ref{eq.spect}) to extract the behavior of degenerate
heavy meson systems in a thermal medium from finite
temperature mesonic correlators. Eq. (\ref{eq.kernel})
is valid only in the continuum. It is not clear in general 
whether $G(\tau, \vec{p})$ calculated on the lattice will satisfy
the same spectral representation, but it was shown in Ref. \cite{karsch3}
that this is the case for the free theory.

\section{DETAILS OF THE LATTICE AND SIMULATION PARAMETERS}
\label{sec.lattices}

In this work we restrict ourselves to the quenched
approximation and use pure gauge lattices generated using
the isotropic Wilson action. In order to have enough points
in the temporal direction at high temperatures, we need
very fine lattices. We use lattices at three different lattice
spacings in the range $ \approx 0.02 - 0.05$ fm. At the
higher temperatures only the finer lattices are used, while
at the lower temperatures, comparison between results from
our finer and coarser lattices gives an idea of the effect
of the limited number of data points in our analysis. The
lattices were generated using the heat-bath and
over-relaxation algorithm, with each sweep consisting of one
heat-bath step followed by 4 over-relaxation steps.
The configurations were separated by 200 - 800 such sweeps, 
the separation in each case being 5-8 times the
autocorrelation time of the Polyakov loop. The exact details of
the lattices are given in Table \ref{tbl.lattices}.
A subset of these lattices were used for the light meson
studies of Ref. \cite{karsch01}.

\begin{table}[htb]
\caption{Lattice parameters. The lattice spacing is
obtained from the string tension.}
\label{tbl.lattices}
\begin{tabular}{ccccccc} 
\hline
$\beta$ & $a^{-1}$(GeV) & $c_{SW}$ & $\kc$ & Size & $T / T_c$ & $ \# $conf \\
\hline
6.499 & 4.04 & 1.494176 & 0.13558 & $48^3 \times 24$ & 0.62 & 50 \\
& & & &  $48^3 \times 16$ & 0.93 & 50 \\
& & & & $48^3 \times 12$ & 1.24 & 50 \\
& & & & $48^3 \times 10$ & 1.49 & 45 \\
6.64 & 4.86 & 1.457898 & 0.13495 & $48^3 \times 24$ & 0.75 & 100 \\
 & & & & $48^3 \times 16$ & 1.12 & 50 \\
 & & & & $48^3 \times 12$ & 1.5 & 60 \\
7.192 & 9.72 & 1.35500 & 0.13437 & $40^3 \times 40$ & 0.9 & 85 \\
 & & & & $64^3 \times 24$ & 1.5 & 80 \\
 & & & & $48^3 \times 16$ & 2.25 & 100 \\
 & & & & $48^3 \times 12$ & 3.0 & 90  \\
\hline
\end{tabular} \end{table}

For the valence quarks, we use the clover-improved \cite{clover}
Wilson action, taking the clover coefficients from the
non-perturbative estimates of the ALPHA
collaboration\cite{alpha}. The clover coefficients used,
and the critical coupling $\kc$ at each $\bt$, are shown in
Table \ref{tbl.lattices}. 
For our coarsest lattice, we use three $\kp$ values that bracket the charm, 
giving a pseudoscalar mass in the range 1.7 - 4 GeV. This allows us to study 
the mass dependence of the properties that we are considering.
At the finer lattices, we use one $\kp$ value close to the charm mass. 

In order to study the lowest states in each of the four channels, we
take operators with four different spin structure. The operators we 
study are listed below, along with the names and spectroscopic 
representations of the lowest states for each channel:
\beq
J_H  = \qquad \begin{array}{lcl}
\bar{c} c & \quad {}^3P_0 \quad & \as \\
\bar{c} \gamma_5 c & {}^1S_0 & \ec \\
\bar{c} \gamma_\mu c & {}^3S_1 & \jpsi \\
\bar{c} \gamma_\mu \gamma_5 c & {}^3P_1 & \ax 
\end{array}
\label{eq.channels} \eeq
For zero temperature spectrum studies, smeared operators are usually 
used to increase the overlap with the ground state, so one can 
study the ground state properties already from correlators at small 
distance. However, when one wants to study other
properties of a channel, for example meson decay constants, one has
to be careful about using smeared operators. Here since we
will be interested mostly in the existence versus
dissolution of the states, it is tricky to use smeared
operators. Smearing mimics bound states even when there
are none \cite{wetzorke01}, and it is also difficult to extract
phenomenological quantities, like the dilepton rate in 
Eq. (\ref{eq.dilepton}), if one works with smeared operators.
Therefore we use point-to-point correlators in
this study. When one knows that there are bound states, it may
be better to use the smeared operators to calculate the
mass of the states. 

The use of the non-perturbative clover action removes
${\cal O}(a)$ discretization errors. However in the case
of heavy quarks discretization errors of order ${\cal O}(a m)$
can be large. The errors in using the clover 
quarks due to finite mass of the quarks 
have been discussed in detail in Ref. \cite{fermilab}. The
main sources of $O(ma)$ errors are in the difference
between the pole mass and the kinetic mass, and in the
renormalization constant used to connect the lattice
operators to continuum ones. For finite $m a$, 
the pole mass that controls the fall-off of the correlator
differs by $O(ma)$ terms from the physically important kinetic mass that
controls the dispersion relation.
This can already be seen from the free quark dispersion relations
\cite{fermilab}:
\ber
a m_0 &=& {1 \over 2 \kp} - {1\over 2 \kc} \nonumber \\
a \mpl &=& \log(1 + a m_0) \label{eq.mass} \\
a \mkin &=& {am_0 (1 + am_0) (2+am_0) \over 2 + 4 am_0 + m_0^2}
\nonumber \eer
where $m_0$ is the standard definition of the bare mass, 
$\mpl$ governs the fall-off of the free quark correlator and
$\mkin$ is the term governing the quadratic momentum dependence 
in the dispersion relation, and therefore, controlling much of the 
spectrum and other interesting physics in heavy quark systems.
A large mismatch between $\mpl$ and $\mkin$ indicates
that the quark is too heavy to be treated relativistically.
Our lattices are fine enough that in most cases the error
due to the finite mass of the charm quark is small enough.
The hopping parameters we have used for our different lattices are 
listed in Table \ref{tbl.heaviness}. We also list there the tree
level pole and kinetic masses, using Eq. (\ref{eq.mass}).
The tree level mismatch between
$\mpl$ and $\mkin$ is $ < 5 \% $ for all the sets, except the
heaviest quark for set I. To see whether the error is also small 
for the interacting theory, we also looked at the 
pole and kinetic masses of the mesons, where 
the pole mass is obtained from
the zero momentum correlator and the kinetic mass is obtained 
from the dispersion relation 
\beq
E^2 = M_{\rm pole}^2 + {M_{\rm pole} \over M_{\rm kin}} p^2. 
\label{eq.kindef} \eeq
While $M_{\rm kin}$ is quite noisy, we found that 
there is hardly any significant
deviation from the relativistic dispersion relation for the
quark masses used by us, except possibly for set IC \cite{heaviness}. 
We therefore use a relativistic treatment for our quarks in what
follows, and set the masses of the mesons from the pole
masses.

\begin{table}[htb]
\vspace{0cm}
\caption{Hopping parameters used and the pole and kinetic masses for
different sets.}
\label{tbl.heaviness}
\begin{tabular}{cccccc}
\hline
$\bt$ & $\kp$ & $a \mpl$ & $a \mkin$ & $a m_0$ & Tag \\
\hline
      & 0.1325 & 0.0889 & 0.0893 & 0.0860 & I A \\
6.499 & 0.1300 & 0.1584 & 0.1608 & 0.1583 & I B \\
      & 0.1234 & 0.3328 & 0.3531 & 0.3640 & I C \\
\hline
6.64  & 0.1290 & 0.1800 & 0.1835 & 0.1821 & II \\
\hline
7.192 & 0.13114& 0.0940 & 0.0945 & 0.0916 & III \\
\hline
\vspace{-0.8cm}
\end{tabular} \end{table}

The lattice operators are connected to the continuum operators as
\beq
J_H^{\rm Cont} = 2 \kp  Z_H(a,m,\mu=1/a) J_H^{\rm Lat} a^{-3}.
\label{eq.norm}
\eeq
The renormalization factors $Z_H(a,m,\mu=1/a)$ are estimated using
1-loop tadpole improved perturbation theory
\beq
Z_H(a,m,\mu=1/a)=u_0(a) (1+\alpha_V(1/a) \tilde z_H)(1+b_H(a) m a)
\eeq
\beq
b_H(a)=(1+\alpha_V(1/a) \tilde B_H)/u_0.
\eeq
Here $\alpha_V(1/a)$ is the running coupling calculated
from the heavy quark potential at scale $\mu=1/a$ \cite{lepage}.
The one loop coefficients $\tilde z_H$ and $\tilde b_H$
were calculated in Refs. \cite{goeckeler} and \cite{sint}
respectively. For the scalar and pseudoscalar channels $\tilde z_H$
was evaluated at scale $\mu=1/a$.
We evaluate $u_0$ from the measured plaquette, \[ u_0 = \langle {1 \over 3} 
{\rm Tr} U_{\rm plaq} \rangle ^{1 \over 4}. \] 
The resulting renormalization factors are shown in Table \ref{zfac}.

\begin{table}[htb]
\vspace{0cm}
\caption{Renormalization factors of local operators for
different sets.}
\label{zfac}
\begin{tabular}{cccccc}
\hline
$\bt$ & $\kp$ & $Z_{PS}$ & $Z_{SC}$ & $Z_{VC}$ & $Z_{AX}$ \\
\hline
      & 0.1325 & 0.782 & 0.839 & 0.900 & 0.926 \\
6.499 & 0.1300 & 0.847 & 0.913 & 0.975 & 1.003 \\
      & 0.1234 & 1.032 & 1.124 & 1.188 & 1.221 \\
\hline
6.64 & 0.1290 & 0.881 & 0.948 & 1.007 & 1.034 \\
\hline
7.192 & 0.13114 & 0.839 & 0.886 & 0.936 & 0.957 \\
\hline
\vspace{-0.8cm}
\end{tabular} \end{table}

We close this section with a discussion of the zero 
temperature masses of the charmonia corresponding to 
the $\kp$ values given in table \ref{tbl.heaviness}.
For our coarsest lattices, we use three $\kp$ values that bracket the 
charm quark. For the other lattices, we use one $\kp$ value
each. Since we have not performed calculations on symmetric
(zero temperature) lattices, our estimates of the zero temperature
masses are based on exponential fits to spatial correlators.
In fact, if one considers the spatial correlators 
on lattices below deconfinement, the finite temperature 
lattices can be considered as small
(in one of the directions) zero temperature lattices.   
For the lattices below $T_c$ used in this study 
(see Table \ref{tbl.lattices}), the small extent is 
about 1 fm for the two coarser lattices (sets I and II) 
and about 0.8 fm for the finest lattice (set III). The 
effect of finite volume on the masses of charmonia was studied 
in detail in \cite{taro03} and it was found that even for 
lattices of size $0.75$ fm there are no sizeable finite size
effects. Furthermore, even for the much larger light mesons, 
spatial correlators calculated on lattices at 0.9 $T_c$
have been seen to give good estimates of
zero temperature masses \cite{laermann}.
The masses we get from exponential fits 
are given in Table \ref{tbl.parameters}. 

\begin{table}[htb]
\vspace{0cm}
\caption{Masses (in GeV) of the different charmonium states obtained
from one exponential fits of spatial correlators below $\tc$.}
\label{tbl.parameters}
\begin{tabular}{ccccc}
\hline
Set & \multicolumn{4}{c}{Mass (in GeV)} \\
 
& $\eta_c$ & $J / \psi$ & $\chi_{c 0}$ & $\chi_{c 1}$ \\
\hline
I A & 1.673(8) & 1.806(8) & 2.151(27) & 2.299(44) \\
I B & 2.440(5) & 2.517(10) & 2.874(30) & 3.006(31) \\
I C & 4.151(4) & 4.191(6) & 4.600(44) & 4.670(48) \\
\hline
II & 3.071(6) & 3.141(12) & 3.526(53) & 3.622(117) \\
\hline
III & 3.744(14) & 3.803(10) & 4.442(86) & 4.666(174) \\
\hline
\vspace{-0.8cm}
\end{tabular} \end{table}

For set II, which is very close to the physical value of
the charm quark mass, we get $70 \pm 12$ MeV for
the hyperfine splitting, consistent with the 
analysis of Ref. \cite{taro03}. The mass splitting between
$P$ and $S$ states is also consistent with the 
findings of \cite{taro03}. For our finest lattices 
the splitting between $P$ and $S$ states is largely 
overestimated. This could be due to the limited 
($\sim 0.82$ fm) extent of the lattice.

\section{Meson correlators and spectral functions for $T<T_c$}
\label{sec.belowtc}

We begin our discussion of spectral functions and temperature effects 
on meson properties with an analysis of the meson
correlators below $\tc$. We look at the temporal
correlators, Eq. (\ref{eq.cor}), for the four channels 
in Eq. (\ref{eq.channels}) and for the lattices with $T < \tc$
as specified in Table \ref{tbl.lattices}. For the lattices at $\bt$ =
6.64 and 7.19, we have one set each below $\tc$, the
spatial correlators from which were used in the previous
section for setting the parameters. Here we use the
temporal correlators and investigate the spectral function
at temperatures $T \simlt \tc$. On our coarsest set at
$\bt$ = 6.49, we have lattices corresponding to two different
temperatures below $\tc$. This allows us to check for any
change in meson properties as one approaches $\tc$ from
below. For the most part of this study we use zero momentum projected 
correlators only, and refer to them as $G(\tau)$, and the corresponding 
spectral function as $\sw$ (we also suppress the subscript
H from now on). 

The operators corresponding to the different channels have
been listed in Eq. (\ref{eq.channels}). 
For the vector channel, we take
the trace over all four directions, 
$\sum_{\mu=0}^3 G^{VC}_{\mu \mu} (\tau)$. Due to current conservation, the
time component, $G^{VC}_{00} (\tau)$,  is a constant, and
therefore, contributes only a delta-function at $\om = 0$ to the
spectral function. In the free quark case, this delta
function cancels a similar contribution from the
space-averaged correlator \cite{karsch2}. In the axial
vector channel, we sum over the three spatial
directions, $\sum_{i=1}^3 G^{AV}_{ii} (\tau)$.

At least in principle, one can obtain all the information
about the spectrum in a given channel by inverting
Eq. (\ref{eq.spect}) to calculate $\sw$. 
Since our lattices have only O(10) independent data points in
the temporal direction, however, 
extracting $\sw$ from lattice data for 
$G(\tau)$ is a highly nontrivial problem. 
Considerable progress towards the solution of this problem has
been made in the last couple of years by using Bayesian
methods \cite{asakawa01}. In the next subsection, we explain the
rudiments of the maximum entropy method (MEM), as employed
in our analysis; in the rest of this section, we use this
method to extract the spectral function, and discuss the
results. 

\subsection{Analysis; MEM methods}
\label{sec.mem}

Bayesian techniques for extracting information from
inadequate data depend on providing prior information to
the analysis in some form \cite{jarrel}. 
In the maximum entropy method (MEM) of data 
analysis, the prior is usually introduced in the form of an
`entropy term' by using the positivity of the spectral
function ($\sigma(\omega) \ge 0, \omega>0$) 
to give it a probability interpretation
\cite{jarrel}. Quite general principles lead to the Shannon entropy
\beq
S = \int_0^\infty d \om [ \sw - m(\om) - 
\sigma (\om) \log (\sigma (\om) / m (\om)) ]
\label{eq.entropy}
\eeq
(see Refs. \cite{asakawa01,jarrel} for justifications for the
form of the entropy term).
The entropy depends on an arbitrary function $m(\om)$,
which is called the default model. This function can be
considered as a part of our prior information.
The central problem becomes then to find the 
$\sw$ that maximizes the `free energy'
\beq
F = L - \al S
\label{eq.freeenergy}
\eeq
or equivalently the conditional probability $P[\sigma |DH]$
of having the spectral function $\sigma$ given some data 
D and the prior knowledge $H$ \cite{jarrel}.
Here $L$ is the likelihood function for the data D, i.e., 
$L = \chi^2 /2$, and $\al$ is a real and positive parameter. 

At zero temperatures, MEM has been
successfully applied in spectrum calculations
\cite{nakahara99,asakawa01,wetzorke00,yamazaki01}. At finite
temperatures, where one is limited further by the small
physical extent of the temporal direction, precise
quantitative information is more difficult to obtain. 
Still, useful qualitative information has been obtained about
the shape of the spectrum and the thermal dilepton 
rate \cite{wetzorke01,karsch01,asakawa02}.

We follow here Bryan's algorithm \cite{bryan} for inverting
Eq. (\ref{eq.spect}). This method uses a singular value
decomposition of the kernel, to find the relevant, nonsingular
directions in the inversion process, which are less than
the number of data points. Then the search for solutions is
restricted to this space, which is now a well-defined
problem. Also, in this algorithm one avoids any explicit
dependence on the external parameter $\al$ in
Eq. (\ref{eq.freeenergy}) by integrating over $\al$ with a
conditional probability (see \cite{bryan,asakawa01} for details).

To reconstruct the spectral function the default model 
$m(\om)$ needs to be specified.
Since we want to explore the modification of the ground
state  peaks in the spectral function, 
it is natural to choose the default model to describe only
the high energy part of the spectral function.
Due to asymptotic freedom the spectral function at
very high energies is
expected to approach the case of two free quarks. 
For two free quarks in a mesonic channel $H$, the spectral
function in the continuum is given by \cite{karsch2}
\ber
\sigma_H(\om) &=& {3 \over 8 \pi^2} \om^2 \Theta(\om^2-4 m^2)
\tanh(\om / 4 T) \sqrt{1-4 m^2/\om^2} \nonumber \\
&{}& ~~~~~~ [ a_H + \frac{4 m^2}{\om^2} b_H ]
\label{eq.fullfree} \\
& \approx & {3 \over 8 \pi^2} a_H \om^2 \qquad {\rm for} \qquad
\om \to \infty . \label{eq.free}
\eer
For the scalar, pseudoscalar, vector and axial vector
channels, the constants $a_H$ are 1, 1, 2 and 2,
respectively \cite{karsch2}. 
On the lattice, the spectral functions for the different 
channels have been calculated recently for the 
free theory \cite{karsch3}. They show a considerably different
high energy structure from Eq. (\ref{eq.free}).
Interaction introduces additional modification of the 
large $\om$ behavior of the spectral function \cite{yamazaki01}.
At low temperatures, where the temporal extent is relatively large and
the dependence on the default model is less important,  
we use $m(\om)=m_0 \om^2$ for the default model,
with $m_0=3/(8 \pi^2)$ for scalar and pseudoscalar channels
and $m_0=3/(4 \pi^2)$ for vector and axial vector channels.
Lattice introduces an ultraviolet energy cutoff 
\[ \om_{\rm max} = {2 \over a} \ln (7 + a m_0) \quad \approx 4 a^{-1} \]
for our masses. This is the frequency cutoff we use in our analysis.
Above $\tc$ the use of a more refined form of the default
model, with more accurate information about the high energy
structure of the interacting theory, becomes necessary,
which we discuss in the next subsection.

Use of the full covariance matrix is desirable in the MEM
analysis. However, it has been shown that with limited 
statistics sometimes the small eigenvalues of the 
covariance matrix can be too noisy and it may be better 
to use some smoothening of the smaller eigenvalues
\cite{michael}. In our analysis, we use the full covariance
matrix in most cases; but sometimes the use of the full matrix
gives rise to an unphysical peak at $\om \sim 0$. In such
cases, we use the minimum smoothing that can get rid of
this peak, if it is possible with a small amount of smoothing.

For the spectral density extracted from noisy data with MEM, 
one can assign error bars to the spectral density
integrated over small regions of frequency (see
Ref. \cite{jarrel}): one defines an `integrated spectral
density', 
\[
H(\om, \Delta) = \int^{\om + \Delta}_{\om - \Delta} d \om
\sw
\]  
and then the covariance of this quantity is calculated as
\beq
\langle \delta H^2 \rangle = \int d \om \int d \om^\prime
\sw \sigma(\om^\prime).
\label{eq.error} \eeq
Since we are concerned with the peak structures in the
spectral function, in the following sections we show the 
error bars on a region integrated over the peak width of
each peak, to show the statistical significance of the
peaks.

\subsection{Discussion of the spectral functions}
\label{sec.results}

As mentioned in the previous section, the reliability of
the spectral function reconstructed from the temporal
correlators using MEM depends on the number of data
points. We start the reconstruction of spectral functions
below $\tc$ on our finest lattice: set III. Besides allowing
us to have the maximum resolution, fine-graining of the
lattice also has the advantage that the lattice artifacts
are shifted further away from the ground state peaks, allowing an
accurate extraction of the ground state peaks of interest. 

\begin{figure}[htb]
\begin{center}
\scalebox{0.65}{\includegraphics{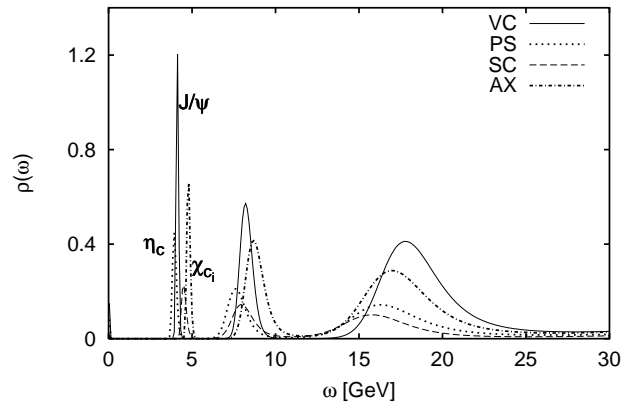}}
\end{center}
\caption{\label{fig.spec0.9} Spectral functions
at 0.9 $T_c$ for the different channels, for set III.}
\end{figure}

Fig. \ref{fig.spec0.9} shows the spectral functions at 0.9
$\tc$, reconstructed from the correlators for this set 
in the different channels. In order
to remove the uninteresting $\om^2$ rise in the spectral
function, here and in everywhere else we plot 
the `reduced spectral function'
\beq
\rho(\om) = \sw / \om^2.
\eeq

At $T=0.9 \tc$ the physical distance available for the
analysis of the temporal correlators is about 0.82 fm, and
we have 40 data points. This allows us to have a good resolution of
the ground state spectral function. We also found that
omitting the data point at smallest $\tau$ enhances the
peak structure for this set. In each channel we find
a three peak structure, with the ground state peak
replicating quite well the properties of the zero
temperature ground state in Table \ref{tbl.parameters}. 
We also find that for energies $\om \sim 2/a$, where the doubler
states are expected to contribute, the structure of
the spectral function is considerably different from the
free spectral function. 
Similar peak like structures have been seen in \cite{yamazaki01}.
As we explain below, we expect the additional peak seen at $\om \sim
1/a$ is also a lattice artifact.

\begin{figure}[htb]
\begin{center}
\scalebox{0.65}{\includegraphics{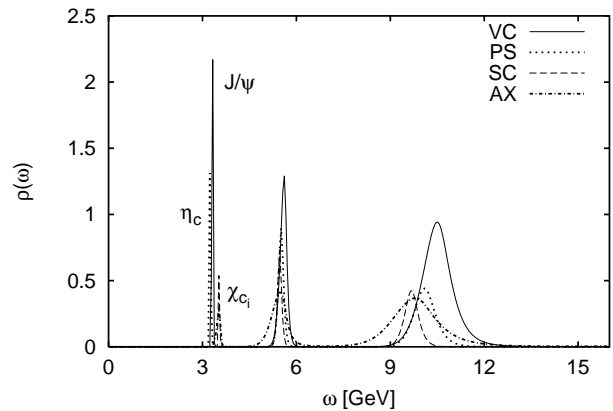}}
\end{center}
\caption{\label{fig.spec0.75b664} Spectral functions
at 0.75 $T_c$ for the different channels, for set II.}
\end{figure}

A comparison with the spectral structure obtained from set
II helps in clarifying the nature of the second peak in
Fig. \ref{fig.spec0.9}. The spectral functions in different
channels for this set are presented in
Fig. \ref{fig.spec0.75b664}. For this figure we use again
the free lattice spectral function as the default model.
The properties of the zero-temperature peak in Table
\ref{tbl.parameters} are reproduced quite well.
The other peaks are seen
to shift at approximately the same ratio from
Fig. \ref{fig.spec0.9} as the inverse lattice spacing. This
confirms that also the second peak obtained in
Figs. \ref{fig.spec0.9}, \ref{fig.spec0.75b664} is dominated by
lattice artifacts. 

In our coarsest set of lattices, at $\bt$ = 6.499, we have
two temperatures below $\tc$. This allows us to study 
the temperature dependence of the mesons below $\tc$, and
will help in introducing certain elements in our analysis,
as we will see below. 

\begin{figure}[htb]
\begin{center}
\scalebox{0.65}{\includegraphics{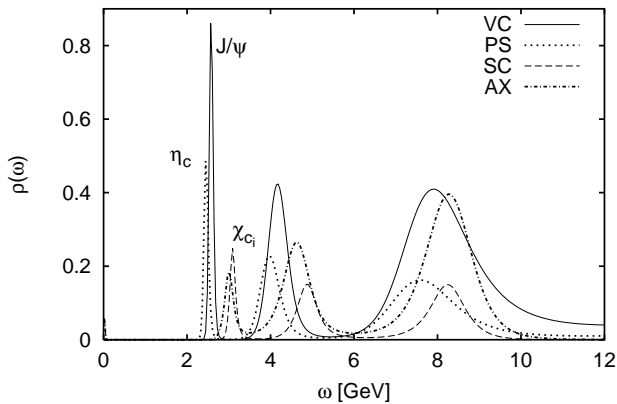}}
\end{center}
\caption{\label{fig.k1300} 
Spectral functions for different channels at $0.6T_c$ 
for set IB.}
\end{figure}

For the correlators at 0.6 $\tc$ the results are shown
for set IB in Fig. \ref{fig.k1300}. We get a three peak
structure for the different channels, similar to the one
shown in the previous figures. The ground state can be
resolved quite reliably, and the strength and position of
the ground state for the different quark masses agree
with the ground state measurements obtained in
previous section.

\begin{figure}[htb]
\begin{center}
\scalebox{0.65}{\includegraphics{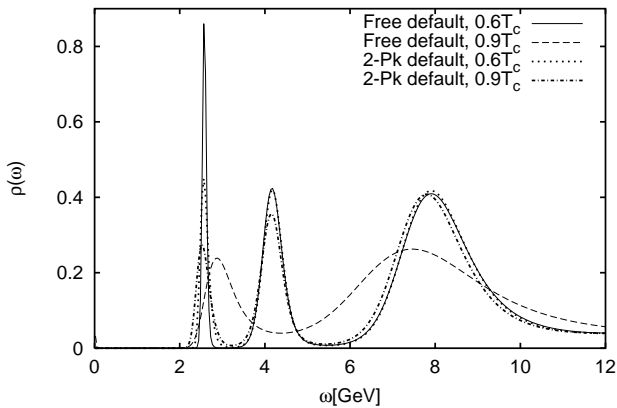}}
\end{center}
\caption{\label{fig.belowtc} Spectral functions for the
vector channel below $\tc$, using both the free spectral 
function and the high energy part of the spectral function
from interacting theory as default model. Set IB.}
\end{figure}

\begin{figure}[htb]
\begin{center}
\scalebox{0.65}{\includegraphics{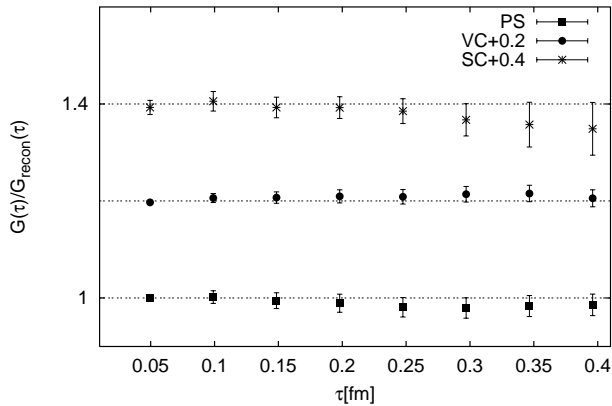}}
\end{center}
\caption{\label{fig.recbelowtc} Ratio of the measured correlators at
0.9 $\tc$ with the reconstructed correlators at this
temperature, using the spectral function at 0.6 $\tc$ (see text),
for different channels and set IB. For visual clarity, the
vector and scalar channels have been shifted horizontally.}
\end{figure}

Figure \ref{fig.belowtc} compares the spectral functions
for the vector channel at 0.6 and 0.9 $\tc$, obtained
with the free spectral function as the default
model. The high $\om$ part of the spectral function at 0.9
$\tc$ is seen to be considerably different from that at 0.6
$\tc$. Also, while $\rhw$ shows a clear ground state peak,
the peak position is slightly changed from that at 0.6
$\tc$ and the fall-off of the peak in the high $\om$ side
is considerably broader. However, we expect this difference
in the spectral function to be an artifact of the limited
number of data points and physical distance at this
temperature. In fact, Fig. \ref{fig.belowtc} does indicate
that at 0.9 $\tc$, the spectral analysis can not resolve
the three peak structure, and the ground state peak is
contaminated by the peak at $\om \sim 1/a$. One way to
improve the reconstruction of the ground state when the
physical distance is small is to provide more accurate
information for the high energy part in the default
model. Since the physical distance covered is small, the
high energy part is still important in the correlator and
it is only when we provide suitable information about the
high energy part that it is possible to extract the
information about the ground state accurately. 
From the analysis of the spectral function at the lowest
temperatures we see that the high energy part of the
spectral function consists of two broad peaks. This is
a general feature of the spectral functions which does not
depend on the default model $m(\om)$ as we will 
see in section \ref{sec.systematics}. So we use a
default model where the high energy part of the spectral
function is taken from the spectral function constructed at
0.6 $\tc$, and in the low energy part we use the model 
\[ m(\om) = m_1 \om^2 \]
with $m_1$ chosen such that the default model is
continuous \cite{lat03}. 
The spectral function constructed at 0.9 $\tc$
using this default model is also shown in
Fig. \ref{fig.belowtc}, where for comparison we also show
the spectral function at 0.6 $\tc$ using the same default
model. As we can see, at 0.6 $\tc$ the spectral function
correctly produces the ground state peak, in both the
position and the strength of the peak. 
(In our present analysis with limited statistics 
the width of the peak is not of direct
physical relevance. It is possibly an effect of the finite
statistics.) Using this default model, now, the
ground state at 0.9 $\tc$ is reconstructed accurately, with
the position and strength of the peak similar to that at
0.6 $\tc$. This also shows the usefulness of using such a
default model, and we will use it for
reconstructions of the spectral function above $\tc$ in the
next section. In essence, what is being done is to realize
that the spectral function for very large $\om >> T$ is dominated by
lattice artifacts, and does not change with temperature.
To reconstruct the ground state accurately from
correlators at short distances it is important to use
correct information of this high energy part.
Another way to confirm that the spectral function does not show 
any significant change between 0.9 $\tc$ and 0.6 $\tc$ 
is by comparing the correlators at 0.9 $\tc$
with the correlators reconstructed from the spectral
function at 0.6 $\tc$ \cite{lat03}. Since the temperature
dependence of the correlators in Eq. (\ref{eq.spect}) comes 
from both the known temperature dependence of the kernel, 
$K(\omega, \tau)$, in Eq. (\ref{eq.kernel}), and the
nontrivial temperature dependence of the spectral function, $\sg$, 
we can try to focus on the temperature dependence of the
spectral function by comparing the correlator with the
reconstructed correlator at that temperature,
\beq
G_{\rm recon, T^*}(\tau, T) = \int d \om \sigma(\om, T^*)
K(\omega, \tau, T) 
\label{eq.recon}
\eeq
where $\sigma(\om, T^*)$ is the spectral function at
another temperature $T^*$. The comparison of the
correlators for the different channels at 0.9 $\tc$ with the 
correlators reconstructed from $\sigma(\om, 0.6 \tc)$ for
the corresponding channels is shown in
Fig. \ref{fig.recbelowtc}. This figure clearly
demonstrates that the correlators at 0.9 $\tc$ are
completely described by the spectral function at the lower
temperature.  

The results from the other quark masses for set I are very
similar and show that for the quark mass range explored by
us, the properties of the ground state mesons do not change
at least up to a temperature of 0.9 $\tc$. This is
consistent with other earlier quenched studies 
in the mesonic sector \cite{laermann} and the
gluonic sector \cite{prd}, which showed that the 
properties of the mesonic
and gluonic states do not change from their vacuum
properties till quite close to the deconfinement transition
temperature. Of course, this result is not unexpected
since one does not expect substantial thermal excitation
of the heavy glueballs
already at these temperatures. Presumably, such a result
may not hold in full QCD, where excitation of pionic states
may produce observable effects already away from $\tc$.

To summarize, in this section we have studied the spectral
functions below $\tc$ for heavy degenerate mesons in
different channels. The ground state properties can be
suitably reproduced by the MEM analysis of the temporal
correlators. We also show that using nontrivial realistic
information about the high energy structure of the spectral
function in the interacting theory allows a more accurate
reconstruction of the ground state already from short
distance correlators available on a small number of data
points. We also propose using the spectral function 
constructed at the lowest available temperature
to reconstruct the correlator at higher temperatures, as a
way of studying the temperature dependence of the
spectral function. This often provides a more robust
analysis, since the lowest available temperature provides
us with the maximum physical distance, allowing for a more
reliable extraction of the spectral function. Both 
these techniques will be particularly relevant in 
the next section for studying the spectral function
above $\tc$, where small physical distance makes the
extraction of the spectral function even more difficult.

\section{Correlators and Spectral Functions for $ T > \tc$}
\label{sec.abovetc}

In this section we present our results for meson correlators 
above $\tc$. We first present results for set II, so that we 
can start with results close to $\tc$, and then go to higher 
temperatures with set III.

A number of interesting and important statements about
system modifications of charmonium spectral functions can
be made by inspecting the finite temperature meson correlators. 
As was discussed in the
previous section, the change in $\sw$ with temperature can
be studied already from the correlators, by first taking
out the trivial temperature dependence of the kernel
$K(\om, \tau, T)$ according to Eq. (\ref{eq.recon}). We
construct $G_{\rm recon}(\tau, T)$ using $\sigma(\om, T^*)$
from the spectral functions obtained for this set at 0.75
$\tc$ (see Fig. \ref{fig.spec0.75b664}). The deviations of 
$\grecon$ from the correlators directly measured above $\tc$ 
then indicate system modifications of the spectral function above 
$\tc$. This method is more robust as it avoids using MEM at the higher
temperatures, where one has less temporal extent and less
data points.

\begin{figure}[htb]
\begin{center}
\scalebox{0.65}{\includegraphics{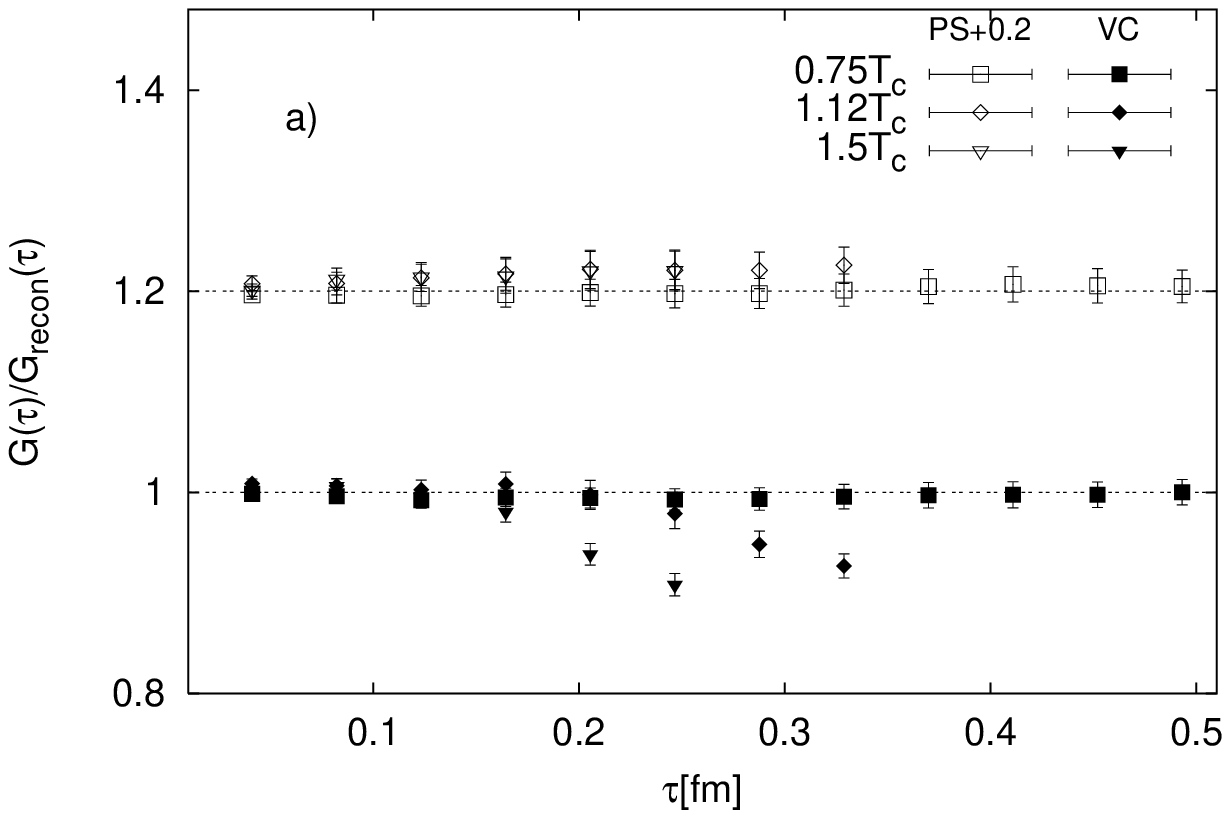}}
\scalebox{0.65}{\includegraphics{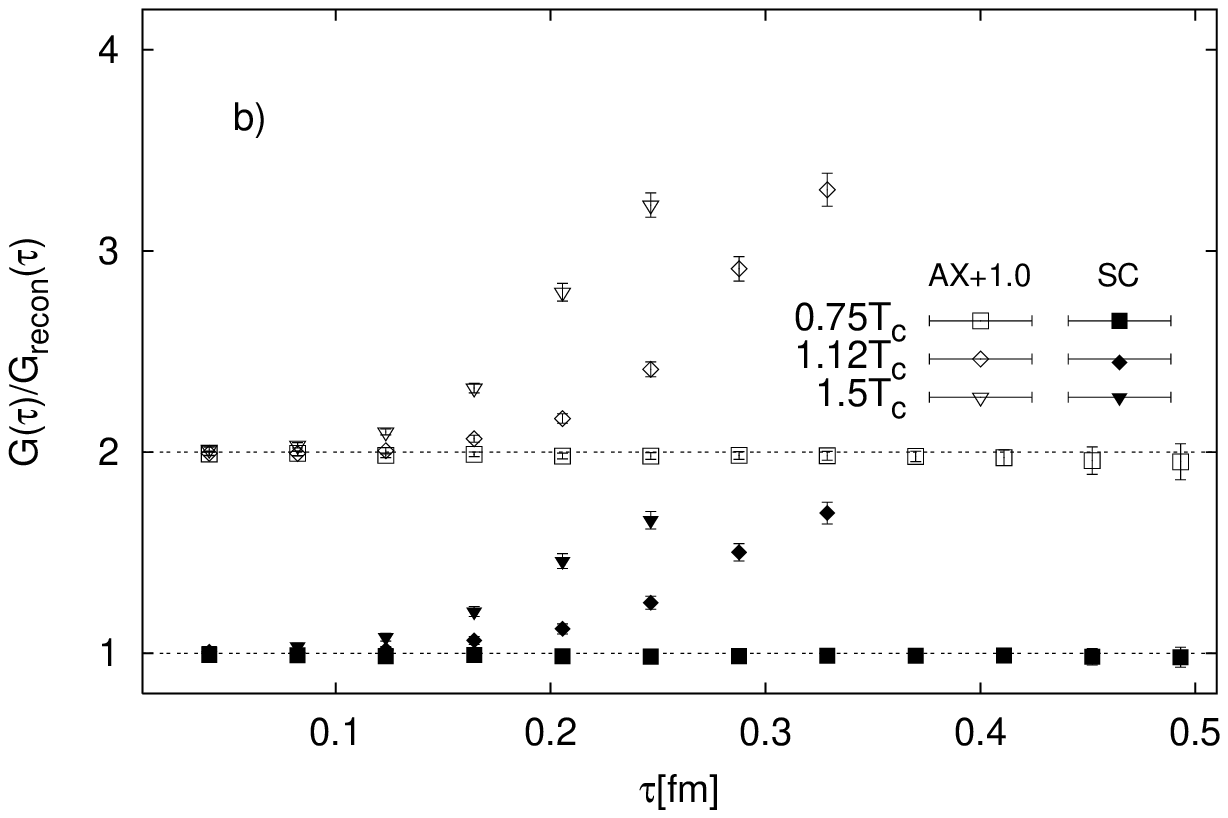}}
\end{center}
\caption{\label{fig.recon664} Ratio of the measured temporal
correlators with the correlators reconstructed from the
spectral function at 0.75 $\tc$ for Set II. a) Pseudoscalar
(PS) and vector (VC) channels; b) Scalar (SC) and axial
vector (AX) channels. For visual clarity, the PS and AX
channel results have been shifted vertically.}
\end{figure}

In Fig. \ref{fig.recon664} a)  we show the results of $\gratio$
for set II for the pseudoscalar and vector channels
(the 1S states). For the pseudoscalar
channel, the ratio is seen to remain equal to one 
upto 1.5 $\tc$, indicating no significant system
modification of its properties upto this temperature. This
is already in sharp contrast to earlier potential model
results, which predicted a dissolution of $\ec$ at
$\approx 1.1 \tc$. For the vector channel, one finds that the
reconstructed correlator explains the data quite well at
small distances. Some significant system modifications,
however, are manifested in the correlators at distances
$>$ 0.25 fm. At 1.5 $\tc$, some modifications are already
seen at distances $\simgt$ 0.2 fm. Later in this section, we
will discuss possible system modifications that can cause
such a change.

The correlators for the scalar and axial vector channels (the 1P states), 
are shown in Fig. \ref{fig.recon664} b). The situation is seen to be quite
different here: already at 1.1 $\tc$, a significant system 
modification of the mesons is manifestated in $\gratio$ 
at all distances. The measured correlators are seen 
to be considerably larger than the reconstructed correlators. 
Such a behavior can be qualitatively understood if the 
1P states are dissolved already at these temperatures: at
low temperatures the spectral function for these states is
zero below $\sim$ 3.5 GeV. If the state is dissolved above
$\tc$, one would expect the correlator to pick up
contributions at significantly smaller $\om \sim 2 m_c$,
making it larger. 

The comparison with reconstructed correlators can only tell
us whether any medium modification can be expected at a
given temperature. The ratios of the correlators give a
rough idea of the magnitude of such modification. To
further explore the nature of the finite temperature
spectral functions it is necessary to reconstruct the
spectral function from the correlators, using MEM.  
Above $\tc$, the physical distance available to
us is small, and fewer data points are available, 
making it difficult to reliably extract
the spectral function without any prior knowledge. 
However, as we have discussed in Sec. \ref{sec.results},    
introducing information about the high energy part of the
spectral function into the MEM analysis in the form of an 
appropriate default model we can overcome this problem.
Therefore, in the default model we use the high energy 
part of the spectral function at 0.75 $\tc$, smoothly
matched to $m_1 \om^2$ in the low energy region (see
Sec. \ref{sec.results}). 

\begin{figure}[htb]
\begin{center}
\scalebox{0.65}{\includegraphics{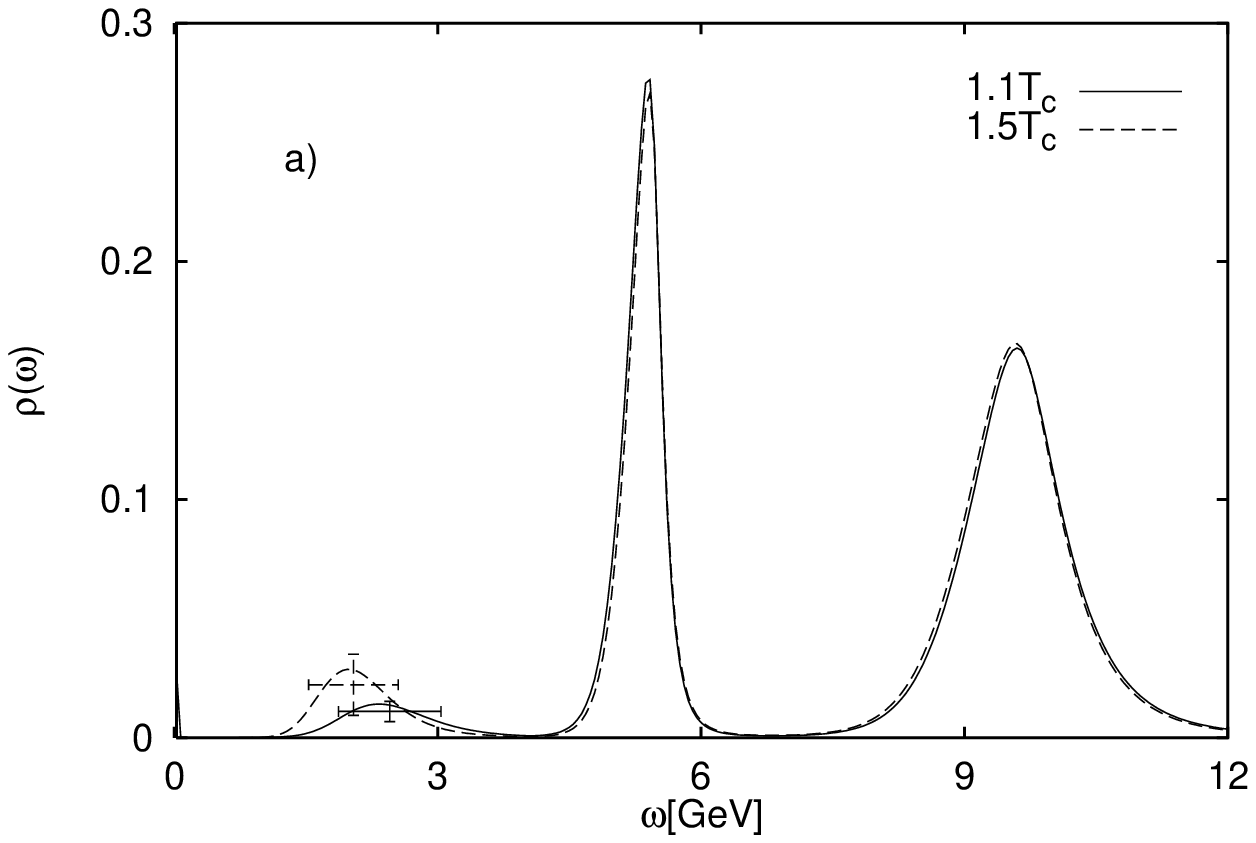}}
\scalebox{0.65}{\includegraphics{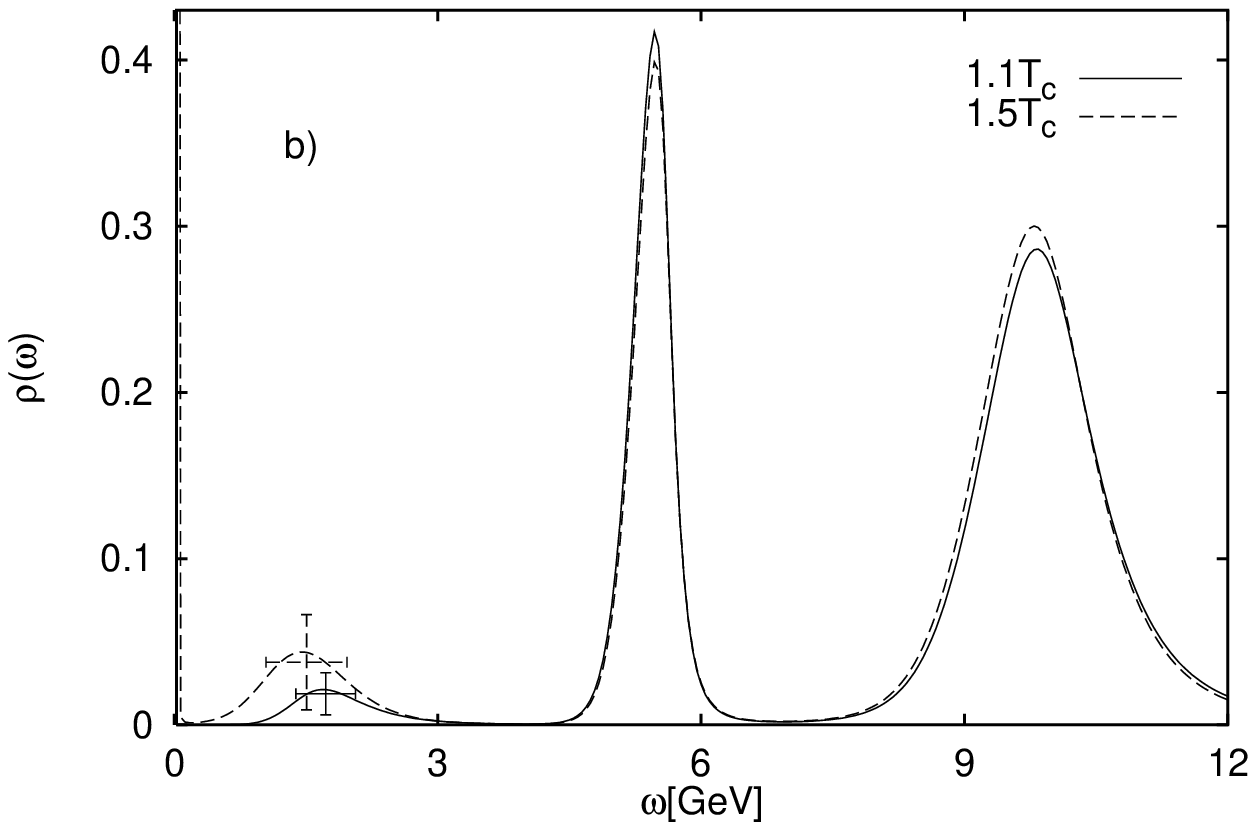}}
\end{center}
\caption{\label{fig.scax664} The spectral function for
the a) scalar and b) axial vector channels constructed from
the temporal correlators, for Set II at temperatures 
between 1.12 and 1.5 $\tc$. The default model uses the high
energy part of the spectral function in
Fig. \ref{fig.spec0.75b664}, as explained in the text.}
\end{figure}

The spectral functions for the scalar and axial vector
channels above $\tc$, reconstructed this way, are shown in
Fig. \ref{fig.scax664}. The significant ground state peak of 
Fig. \ref{fig.spec0.75b664} for these channels are not
found at these temperatures. A non-zero spectral function 
is seen at a significantly lower $\om$, but the peak
structure is statistically not significant, and it could be
related to a branch cut coming from two on-shell propagating 
quarks. Such a $\sw$ at low $\om$
can also explain the rise of the correlators seen in
Fig. \ref{fig.recon664} b). Since the quark mass for set II
is very close to the physical charm quark (see Table
\ref{tbl.parameters}), Fig. \ref{fig.scax664},
together with Fig. \ref{fig.recon664} b), will indicate
that the $\as$ and the $\ax$ are seriously modified,
possibly dissolved, already at 1.1 $\tc$.

\begin{figure}[htb]
\begin{center}
\scalebox{0.65}{\includegraphics{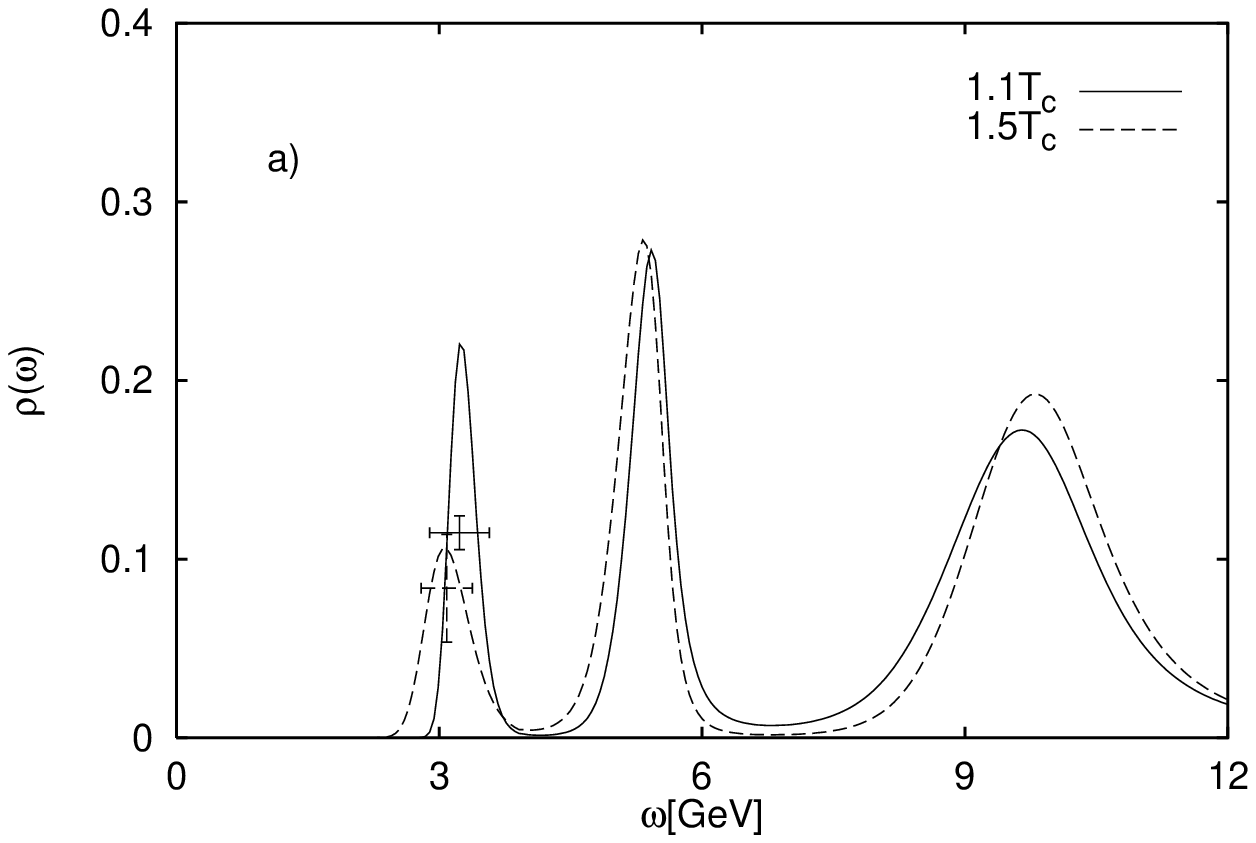}}
\scalebox{0.65}{\includegraphics{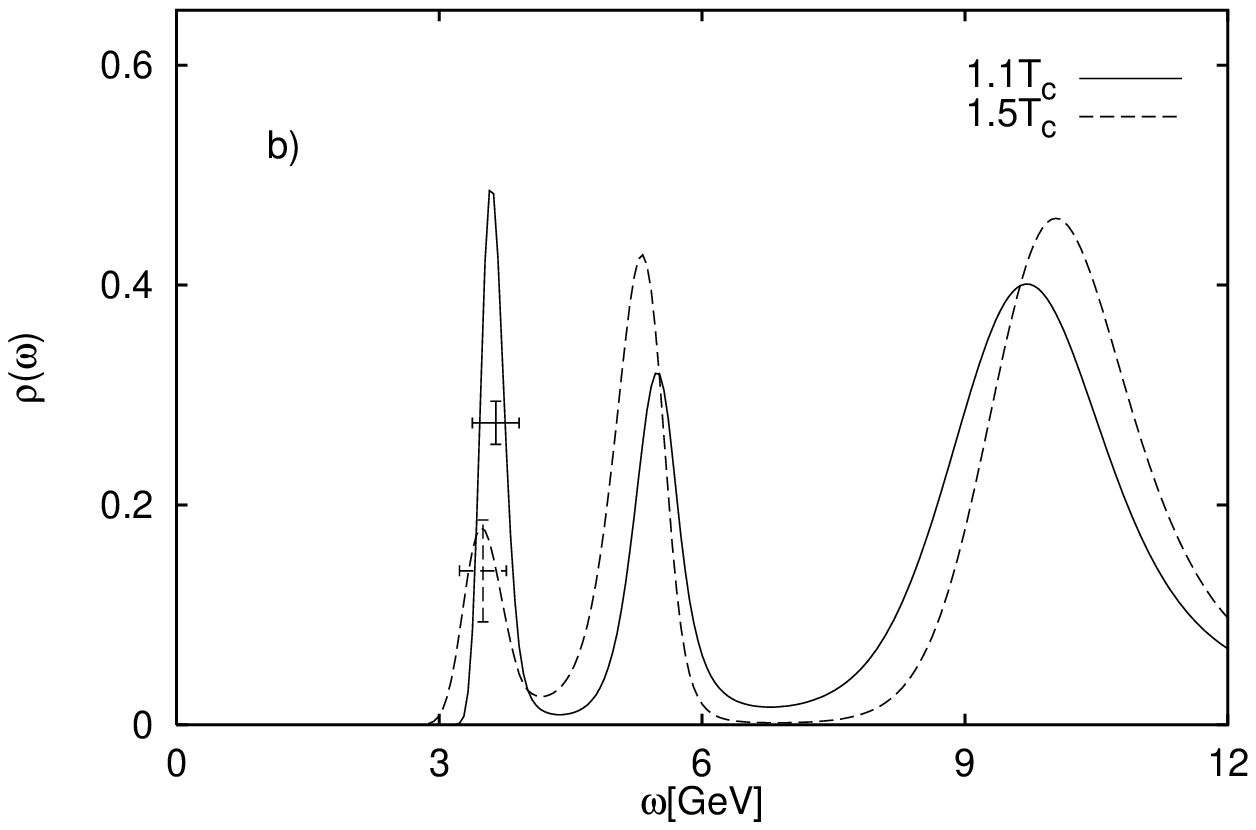}}
\end{center}
\caption{\label{fig.psvc664} The spectral function for the
a) pseudoscalar and b) vector channels constructed from
the temporal correlators, for Set II at temperatures 
between 1.12 and 1.5 $\tc$. The default model uses the high
energy part of the spectral function in
Fig. \ref{fig.spec0.75b664}, as explained in the text.}
\end{figure}

The situation is different for the 1S states: 
Figure \ref{fig.recon664} a) seems to rule out dissolution
or a major modification of these states. The reconstructed
spectral functions, shown in Fig. \ref{fig.psvc664},
support this: a statistically significant ground state peak
is seen upto 1.5 $\tc$, showing that the $\ec$ and the
$\jpsi$ survive at least upto this temperature. While no
major change in mass is indicated, a reduction in peak
strength is seen at 1.5 $\tc$; however, as discussed 
below and also Sec. \ref{sec.systematics}, 
Fig. \ref{fig.664nd6}, we expect it to be
related to the associated systematics.

\begin{figure}[htb]
\begin{center}
\scalebox{0.65}{\includegraphics{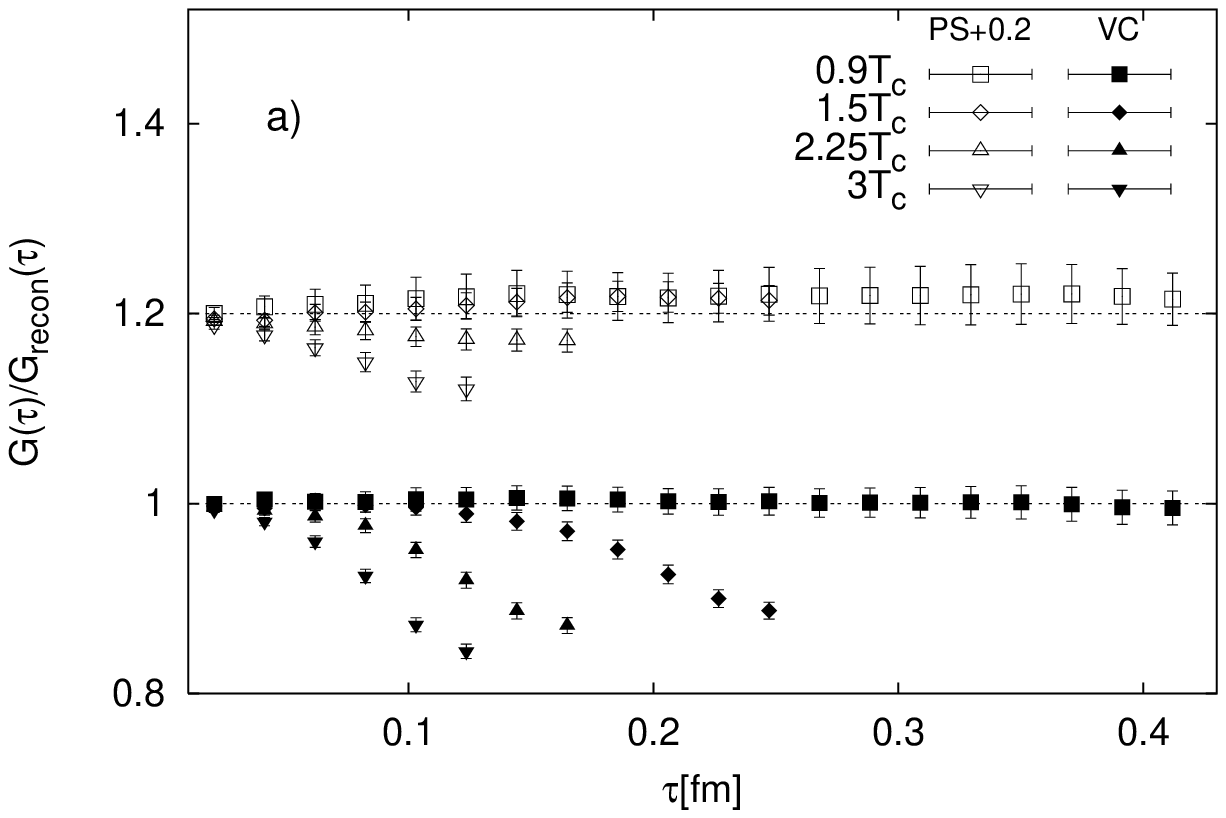}}
\scalebox{0.65}{\includegraphics{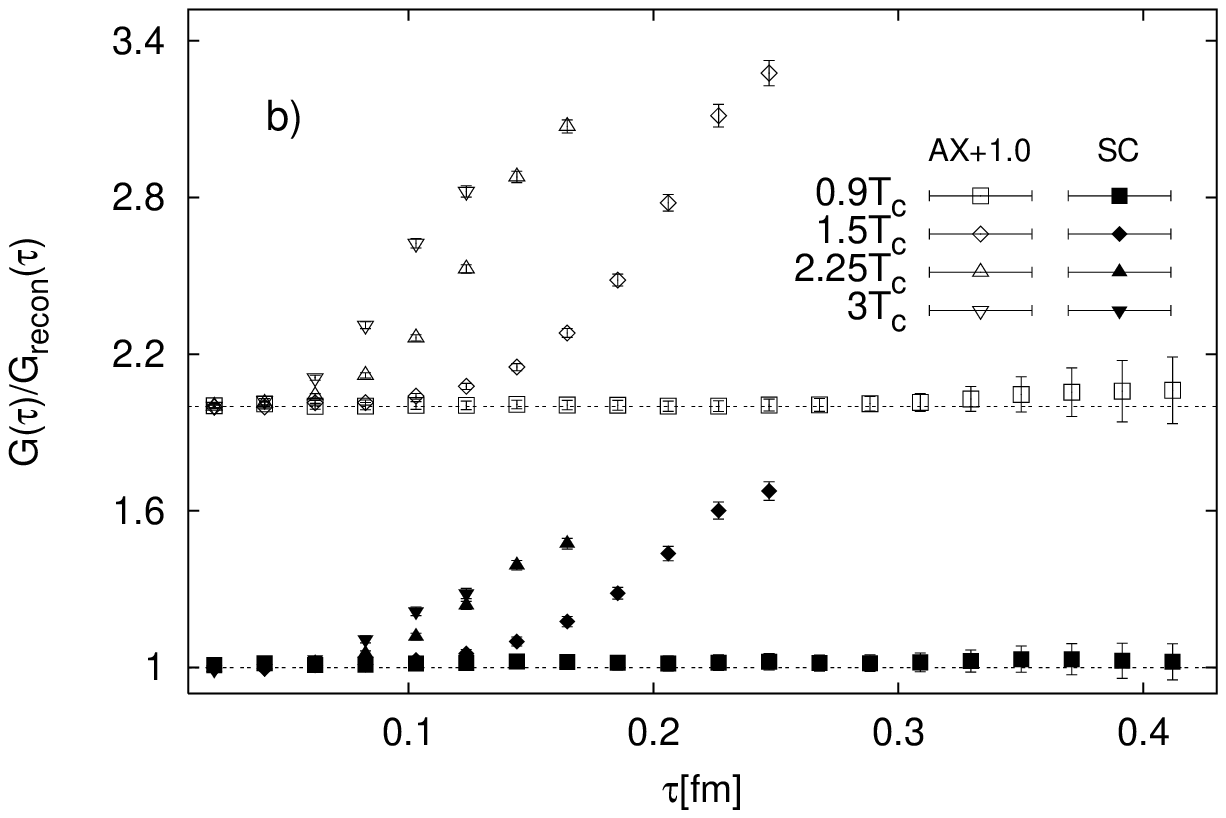}}
\end{center}
\caption{\label{fig.recon719} Ratio of the measured temporal
correlators with the correlators reconstructed from the
spectral function at 0.9 $\tc$ for Set III. a) Pseudoscalar
(PS) and vector (VC) channels; b) Scalar (SC) and axial
vector (AX) channels. For visual clarity, the PS and AX
channel results have been shifted vertically.}
\end{figure}

Having seen that the $\jpsi$ survives till 1.5 $\tc$, it
will be interesting to explore its properties at higher
temperatures. Our finest lattices at set III allow us to 
go upto a temperature of 3 $\tc$. The finer grid at 1.5
$\tc$ here also allows one to check the reliability of the
results of set II. 
Figure \ref{fig.recon719} a) presents $\gratio$ for different
temperatures. $\grecon$ is constructed using the spectral
functions at 0.9 $\tc$ presented in
Fig. \ref{fig.spec0.9}. Upto 1.5 $\tc$, the features of
Fig. \ref{fig.recon719} a) are very similar to that of
Fig. \ref{fig.recon664} a). For the pseudoscalar channel, the
data is completely explained by the spectral function at
0.9 $\tc$. Some small, but statistically significant,
deviations appear at 2.25 $\tc$. At 3 $\tc$, we see
significant modification at all distances. For the vector
channel, at 1.5 $\tc$ the data agree with the reconstructed
correlator at small distances, but start showing
significant modifications at larger distances, $\tau \ge$ 0.15 fm.
At 2.25 $\tc$ such modifications appear already at $\sim 0.1$
fm while at 3 $\tc$, $\gratio$ differ significantly from 1
at all distances.

For the scalar and axial vector channels, one sees at 1.5
$\tc$ a very similar enhancement of the correlator to
Fig. \ref{fig.recon664} b), with $G(\tau, 1.5 \tc)$ $\sim
60 \% $ larger than $\grecon$ at $\sim$ 0.25 fm for the scalar
and $\sim 120 \% $  larger for the axial vector. The pattern
of enhancement continues as one goes to higher temperatures.
 
\begin{figure}[htb]
\begin{center}
\scalebox{0.65}{\includegraphics{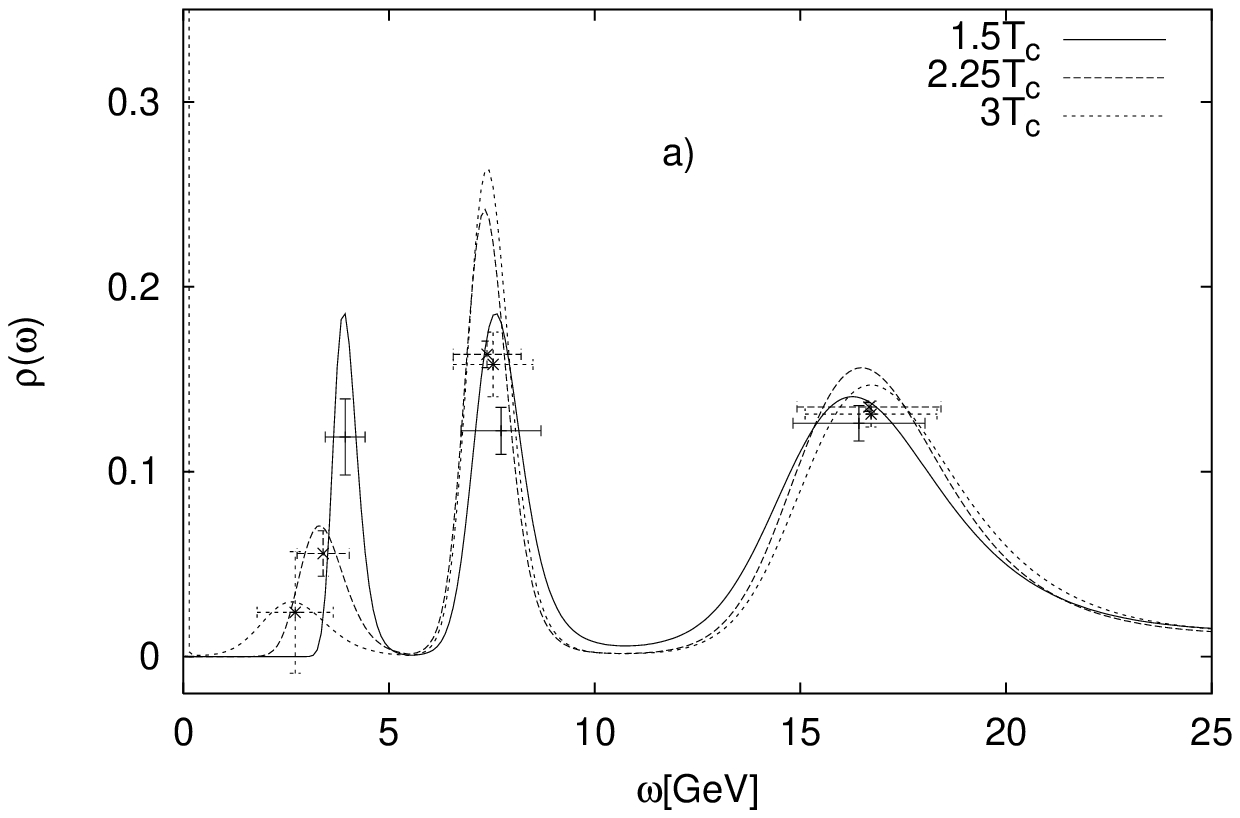}}
\scalebox{0.65}{\includegraphics{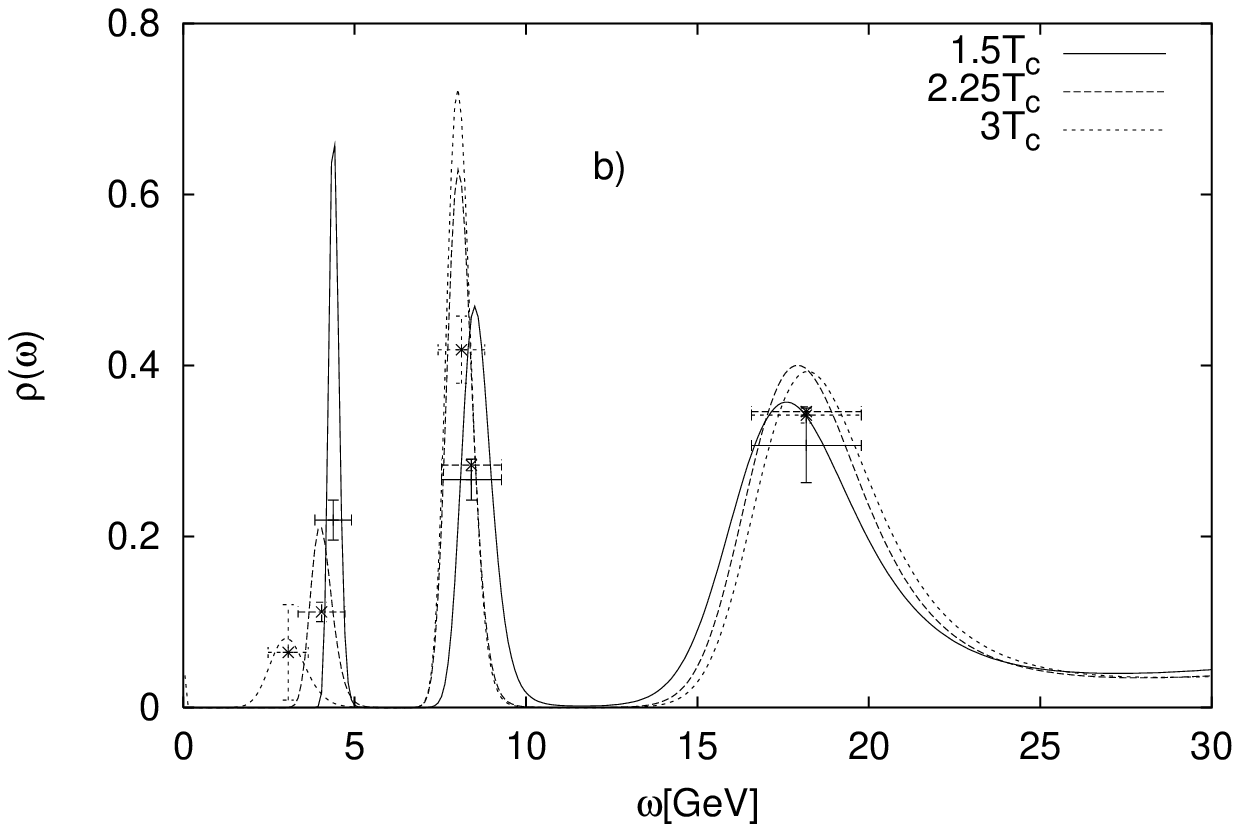}}
\end{center}
\caption{\label{fig.psvc719} The spectral function for
a) pseudoscalar and b) vector channels constructed from
the temporal correlators, for Set III at temperatures 
upto  3 $\tc$. The default model uses the high
energy part of the spectral function in
Fig. \ref{fig.spec0.9}, as explained in the text.}
\end{figure}

To extract the spectral function from the temporal
correlators above $\tc$, we use, as input to the default model,
the large $\om$ structure shown in Fig. \ref{fig.spec0.9}.
Fig. \ref{fig.psvc719} a) shows the spectral
functions above $\tc$ obtained with this default model. A
strong and statistically significant ground state peak is
obtained at 1.5 $\tc$. The peak position and peak strength
are found to be very similar to those at 0.9 $\tc$,
indicating essentially no significant change for the
pseudoscalar upto this temperature. This is completely
consistent with the trend seen for $\gratio$ for the
pseudoscalar. At 2.25 $\tc$, we still see a statistically
significant peak, but with a reduced strength, and at 3
$\tc$ the structure at the peak position is very weak and 
not statistically significant. 

For the vector channel, the temperature dependence of
the spectral function is very similar. A very strong peak
is seen at 1.5 $\tc$, with no observed reduction in the
peak strength from that below $\tc$. Unlike the
pseudoscalar channel, we see here a slight shift of the
peak position, but the amount of the shift depends on the
systematics and we are unable to make a definite comment on
it. At 2.25 $\tc$, a significant peak is seen with a
reduced strength, and at 3 $\tc$, the structure at the peak
position is too broad to be interpreted as a resonance, and
is also statistically not significant.

As we will discuss in detail in the next section, the 
details of the features presented in Fig. \ref{fig.psvc719}
and Fig. \ref{fig.psvc664} are not dependent on the
systematics incorporated in extracting the spectral
function, either from the specifics of the default model or
from the small temporal extent and limited number of data
points at higher temperatures. The 1S state mesons survive
with no significant change of strength at least upto 1.5 $\tc$, 
and at 2.25 $\tc$ a significant peak still survives but
with a reduced strength, while at 3 $\tc$ system effects
reduce the peak strength enough that one should consider
the state to be dissolved. Such a change in the spectral
function is also completely consistent with the behavior of
$\gratio$ for the pseudoscalar channel, since a reduced ground
state peak will cause a depletion of $G(\tau)$. For the
vector channel, the general picture indicated by
Fig. \ref{fig.psvc719} b) is the same. However, we also see
here a depletion of $G(\tau)$ at large distances already at
1.5 $\tc$, but no reduction of the peak strength in
Fig. \ref{fig.psvc719} b), and a small increase in mass. 
While such an increase in mass at 1.5 $\tc$ is consistent
with the depletion of the correlator, we cannot
isolate any physical mass shift from the systematics.

We would like to comment here on similar works on $\jpsi$
in the quark-gluon plasma. Ref. \cite{umeda02} uses smeared
operators, which allows them to study the properties of the
in-medium $\ec$ and $\jpsi$ in more detail, and finds a large Breit-Wigner
width $\sim$ 120 MeV and 210 MeV, respectively, at 1.1
$\tc$. They also find no reduction of the masses of the 1S states
above $\tc$. With our point-point correlators we cannot extract a
width reliably; however, any such width will cause an
enhancement of the correlator, which we do not see for
these states. However, it is possible that such an
enhancement is shielded by a corresponding reduction in
peak strength. Ref. \cite{asakawa03} uses point-point correlator, and
concludes that the 1S states dissociate already at 1.9
$\tc$. Of course, our quarks for set III are a little
heavier than physical charm, and that can cause the
mismatch (the mass dependence of the modification pattern
is discussed below); however, we think another likely reason could be
that due to the use of the free continuum spectral function, 
Eq. (\ref{eq.fullfree}), as the default model in Ref. \cite{asakawa03},
(which does not provide accurate information about the structure of the 
high energy part of the spectral function),
and the small temporal extent at this temperature, 
MEM cannot resolve the peak structure correctly. 

For our coarsest lattices, set I, we have three quark masses, giving
pseudoscalar masses in the range 1.7 to 4.1 GeV (see Table
\ref{tbl.parameters}). This allows us to study the mass
dependence of the dissolution pattern, i.e., to answer the
question whether the ``melting temperature'' for the quarkonia
has a mass dependence. 

\begin{figure}[htb]
\begin{center}
\scalebox{0.65}{\includegraphics{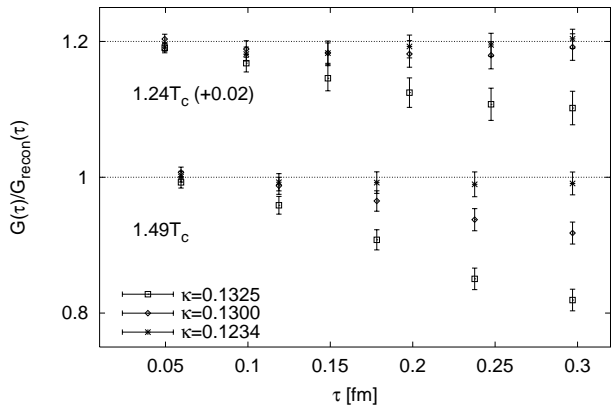}}
\end{center}
\caption{\label{fig.rat649} The mass dependence of $\gratio$
at different temperatures, for the pseudoscalar channel.
The refernce spectral function used for $\grecon$ is 
$\sigma(\om, 0.62 T_c)$. For clarity, the points at 
1.24 $\tc$ have been shifted vertically, as indicated in the figure.}
\end{figure}

In Fig. \ref{fig.rat649} we show the ratio of the
pseudoscalar correlators at different temperatures
with the correlators reconstructed from spectral
function at 0.62 $\tc$ using Eq. (\ref{eq.recon}). At 0.93
$\tc$, expectedly, the correlators are explained by the
spectral function at 0.62 $\tc$ for all quark masses. For
the lightest quark studied by us ($\kp$ = 0.1325) some
system modification is already seen at 1.24 $\tc$, and
the modification becomes larger as one goes to 1.5
$\tc$. For the next heavier quark ($\kp$=0.13), at 1.24 $\tc$ the
correlators agree completely with the reconstruted
correlators, and one sees temperature modification of the
pseudoscalar only at the higher temperature. For our
heaviest quark ($\kp$=0.1234), 
Fig. \ref{fig.rat649} a) shows that the
pseudoscalar correlator is completely explained by the
spectral function upto at least 1.5 $\tc$. The vector
channel shows a similar trend, with the lighter quarks showing
larger medium modification at a given temperature.
For the scalar and axial vector channels, we found
significant medium modifications similar to
Figs. \ref{fig.recon664} b) and \ref{fig.recon719} b) for
all three quark masses, indicating that also for the
heaviest quarkonia studied by us, the 1P states undergo 
significant medium modification, possibly dissolution,
already at 1.24 $\tc$.

\section{Discussion of possible systematic errors in MEM
analysis}
\label{sec.systematics}

The extraction of the continuous spectral function from O(10) data 
points is, 
to start with, a rather delicate problem, and one needs to
understand the effect of the systematics well before
drawing physics conclusions. In this section we are going 
to discuss possible systematic effects in the spectral
functions presented in sections \ref{sec.results} and
\ref{sec.abovetc}. One obvious problem is that at high temperature
both the number of data points and physical extent of the temporal
direction becomes small, which make the reconstruction of
the spectral functions difficult \cite{umeda02}.
To some degree this problem can be overcome if one uses the information about
the high energy behavior of the spectral function extracted at low
temperatures, as was discussed in section IV. This issue will
be discussed here in more detail.

\begin{figure}[htb]
\scalebox{0.65}{\includegraphics{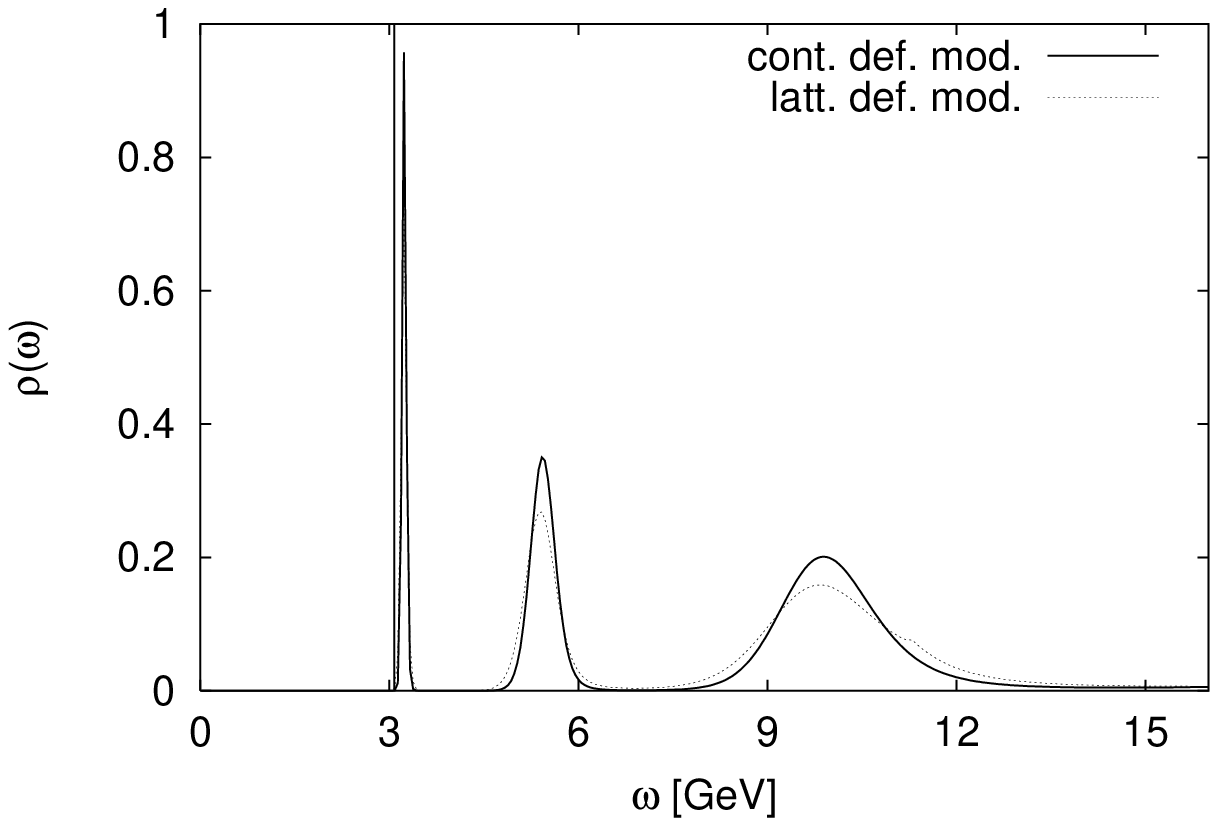}}
\scalebox{0.65}{\includegraphics{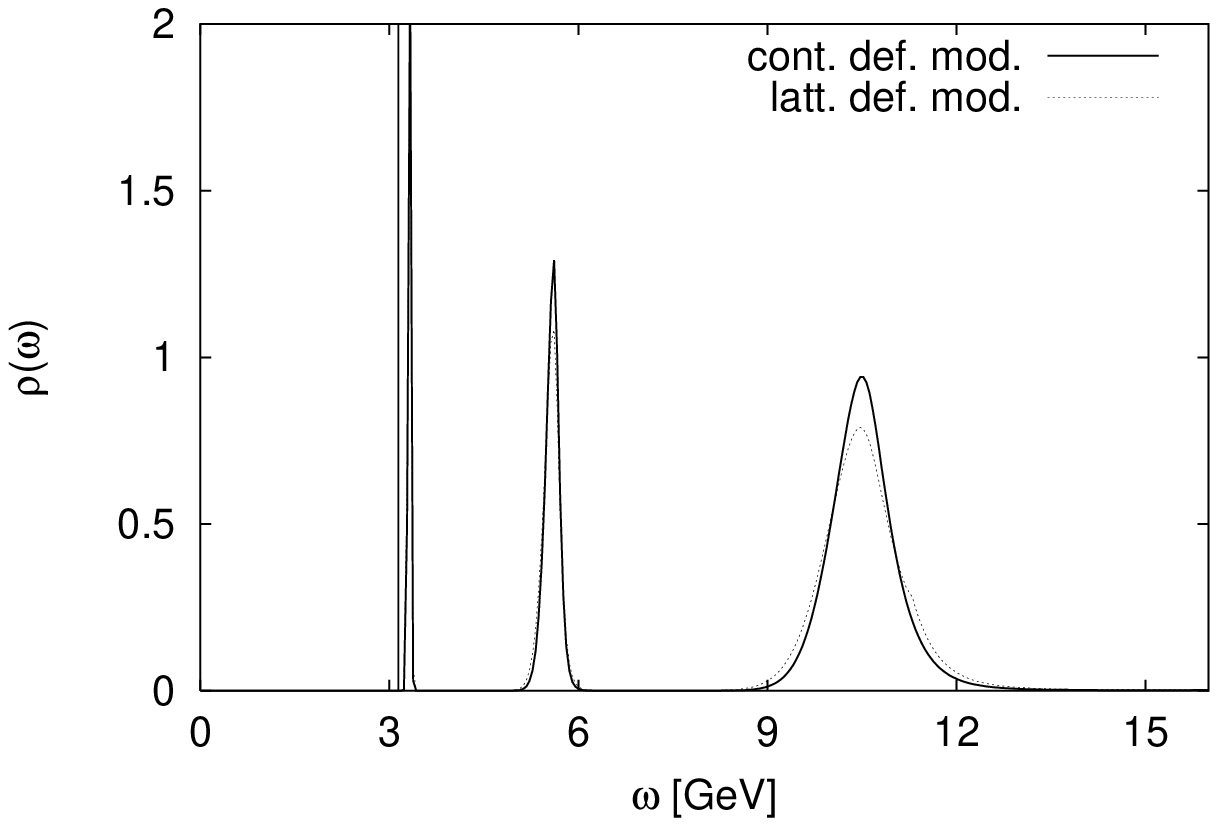}}
\caption{\label{fig.664syst} 
The pseudoscalar (top) and vector (bottom) spectral
functions at $0.75T_c$ reconstructed using the lattice
and continuum default model (see text for further details).
}
\end{figure}

We start the discussion of the systematic effects for spectral functions
corresponding to ground state, i.e. pseudoscalar and vector ones.
In our method of analysis, the spectral function below
$\tc$ plays a crucial role since we extract the high energy
information from it, and we examine first the default model
dependence of this spectral function. 
In Fig. \ref{fig.664syst} we compare the pseudoscalar 
and vector spectral functions at $0.75T_c$ 
reconstructed using the free massive lattice spectral
function as a default model, with those shown in 
section IV where the massless continuum spectral function
was used as a default model. For the quark mass in the free
lattice model we use $m a
\approx 0.17$, which corresponds to the bare mass for this
set. As one can see from the figure the dependence on the 
default model is quite small, even though the two default
models have very different structures. For comparison, 
the masses extracted from the spatial correlators at this
temperature are also shown as vertical lines.
There is a small deviation between the peak position and the 
screening masses, $\sim 3\%$ for the pseudoscalar 
channel and $\sim 7\%$ for the vector channel.
Since we do not expect medium modification of 
quarkonia properties below deconfinement (cf. Fig. \ref{fig.recbelowtc})
this should be interpreted as systematic error in our MEM
analysis. Similar statements also hold for the spectral 
functions at $0.9T_c$. 

\begin{figure}[htb]
\scalebox{0.65}{\includegraphics{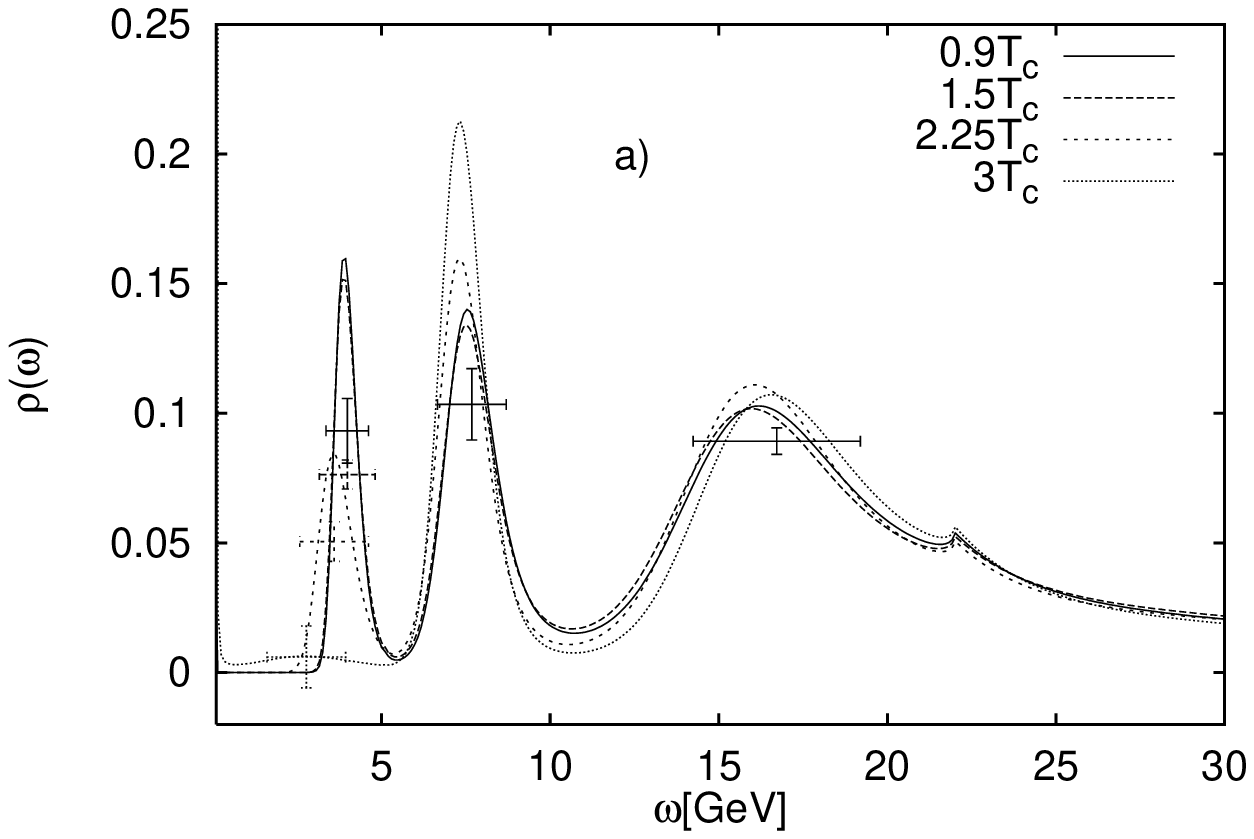}}
\scalebox{0.65}{\includegraphics{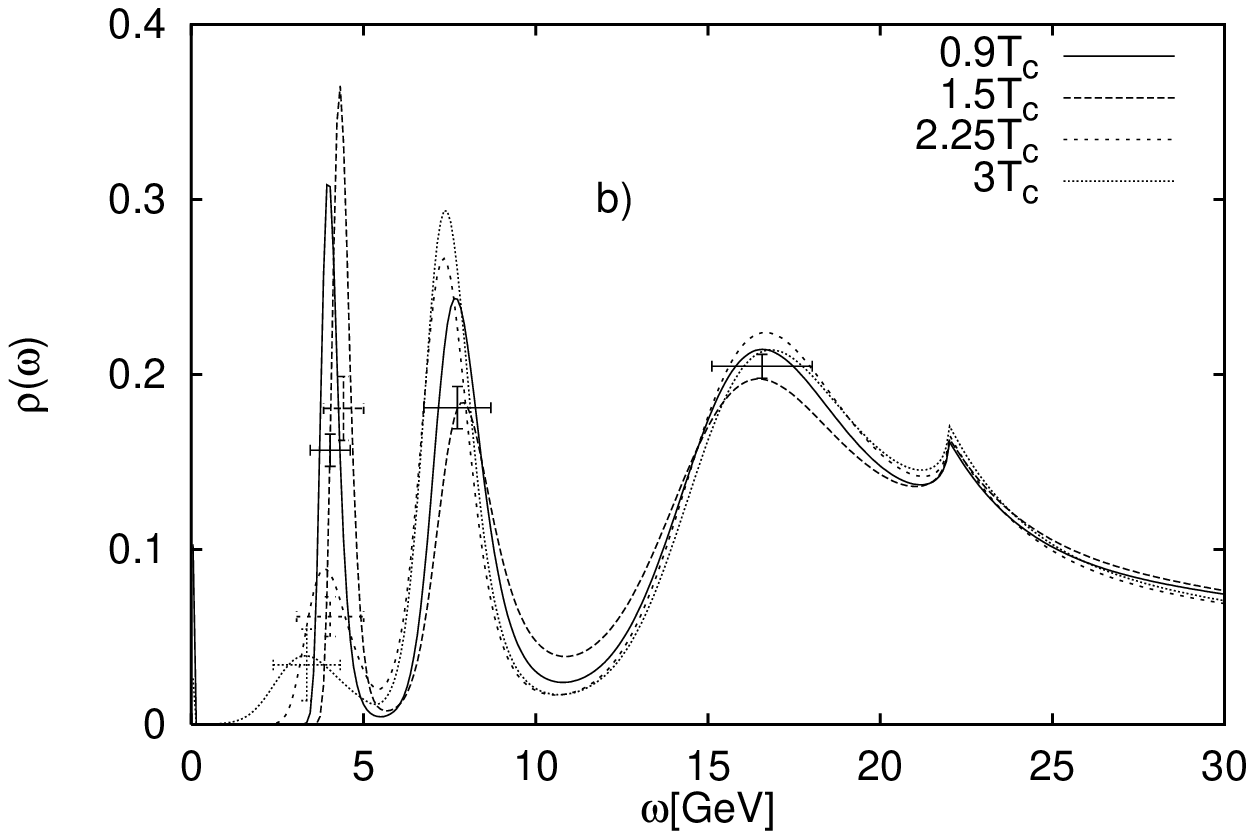}}
\caption{\label{fig.lat719} The spectral functions 
for a) pseudoscalar and b) vector channels at different
temperatures for set III, where the default model incorporates the high
energy part of the spectral function at 0.9 $\tc$ obtained
using the free massive lattice spectral function (see text
for further details).}
\end{figure}

Therefore, our way of using the information from the high $\om$ part of the
spectral function below $\tc$ in the default model for the
analysis above $\tc$ removes the arbitrariness of the
default model. To check this, in Fig. \ref{fig.lat719}
we redo the analysis presented in Fig. \ref{fig.psvc719},
where now the high $\om$ part was taken from the spectral
function at 0.9 $\tc$ calculated with the free massive
lattice default model with the corresponding mass. For
comparison, we also show the spectral function at 0.9 $\tc$
recalculated with this default model. As one can see, the
main features of Figs. \ref{fig.lat719} a) and b) are
very similar to those of Fig. \ref{fig.psvc719} a) and b),
respectively, so that one can essentially substitute them 
without having to change any of the discussion following 
Fig. \ref{fig.psvc719}.

\begin{figure}[htb]
\scalebox{0.65}{\includegraphics{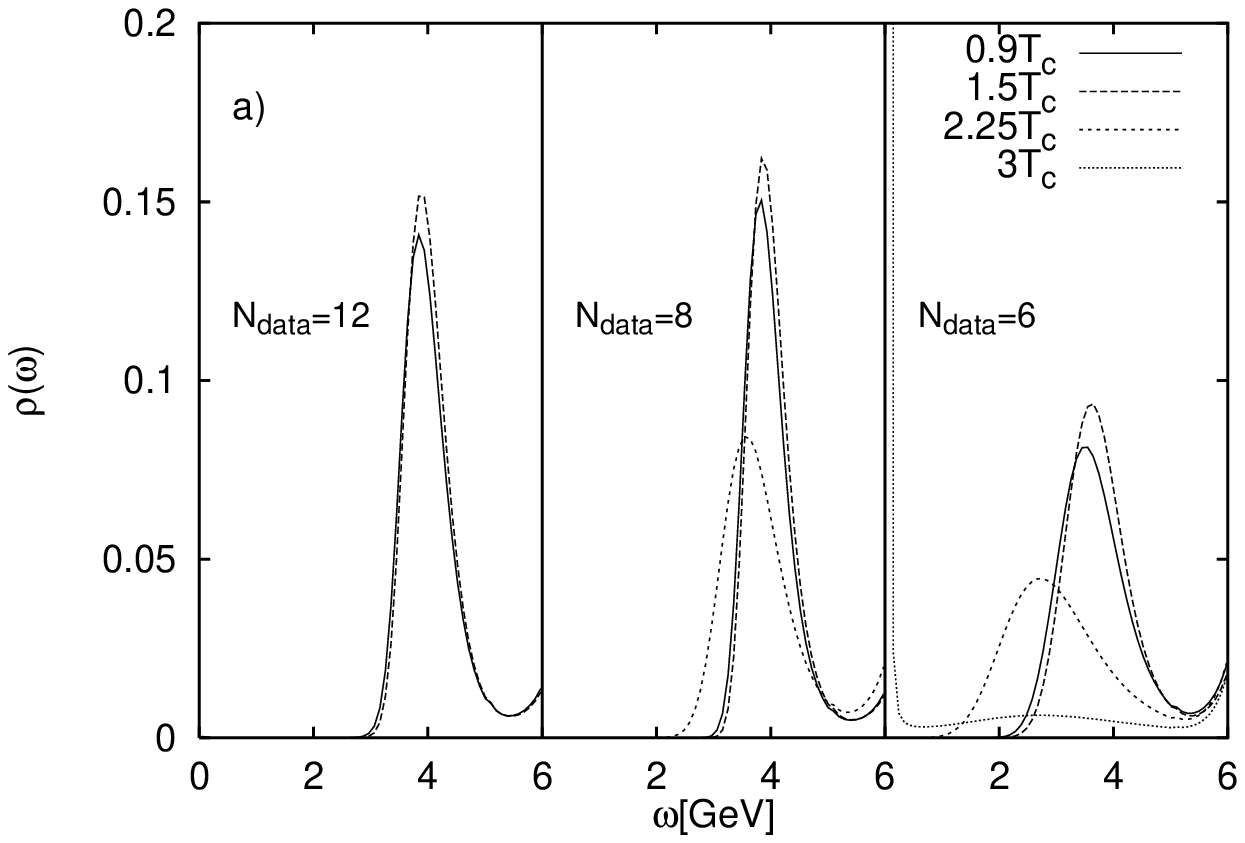}}
\scalebox{0.65}{\includegraphics{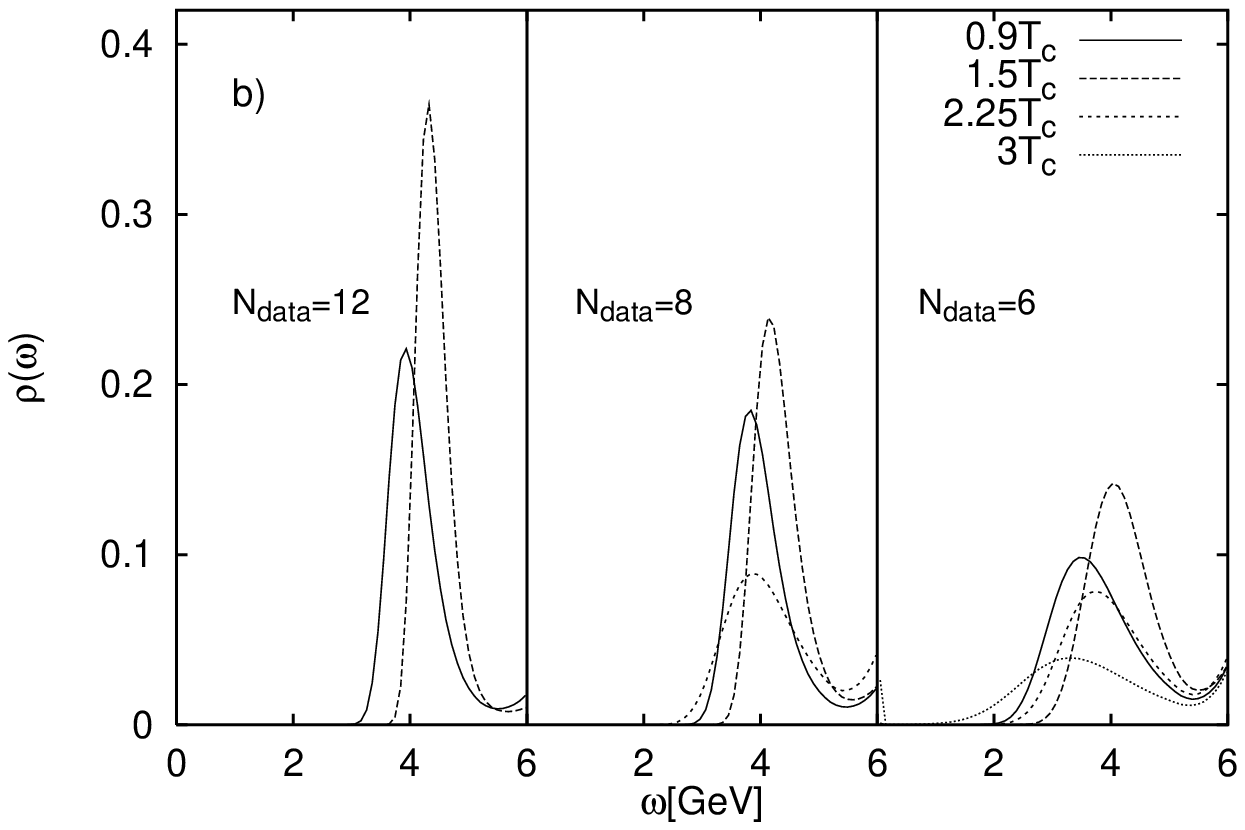}}
\caption{\label{fig.719nd} The a) pseudoscalar and 
b) vector spectral functions at different 
temperatures reconstructed using the same number of 
data points at each temperature.}
\end{figure}

Another question to be addressed is to what extent the
temperature modification of the spectral functions seen in
Sec. \ref{sec.abovetc} and Fig. \ref{fig.lat719} are real 
physical effects, as opposed to the offshoot of the
inability of the MEM to correctly reproduce the spectral
function from the small number of data
points and small temporal extent available at higher temperatures.
To answer this question we compare in Fig. \ref{fig.719nd} the
spectral functions for set III at different temperatures
constructed using the same number of data points. In
the three panels of Fig. \ref{fig.719nd} a) the
pseudoscalar spectral functions at different temperatures
are constructed with the same number of data points as are
available for the highest temperature in the panel,
omitting points from the center for the lower temperatures. 
${\rm N}_{\rm data}$ here refers to the number of points
{\cal after} periodic folding. Though the figure shows
that the details of reconstructed spectral functions depend
on the number of data points used in the analysis, it is
clear that the fact that substantial modification of the spectral 
function takes place at $2.25T_c$ and the ground state is 
"dissolved" at $3T_c$ are not artifacts of the limited resolution.  
Fig. \ref{fig.719nd} a) also show very clearly the absence
of any significant change in the pseudoscalar channel upto
1.5 $\tc$. Similar comments also hold for the vector
channel in Fig. \ref{fig.719nd} b). In this case, we do see
the mass shift at 1.5 $\tc$ discussed before irrespective
of ${\rm N}_{\rm data}$, but the exact amount of the shift
differs. 

\begin{figure}[htb]
\scalebox{0.65}{\includegraphics{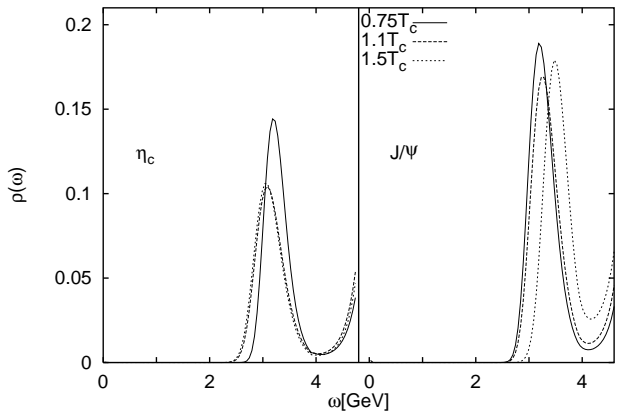}}
\caption{\label{fig.664nd6} Spectral functions for set II for 
pseudoscalar and vector with ${\rm N}_{\rm data}=6$ data points.}
\end{figure}

In Fig. \ref{fig.664nd6} we do a similar analysis for set
II, where we reconstruct the spectral function at different
temperatures using the same physical extent as available
for 1.5 $\tc$. This figure shows that the seeming reduction
of the peak strength at 1.5 $\tc$ in
Fig. \ref{fig.psvc664} was probably due to the small
number of data points available there, and the pseudoscalar
and vector peaks are not substantially changed upto 1.5
$\tc$, as one would expect from Fig. \ref{fig.recon664} a)
and the results for set III.

\begin{figure}[htb]
\scalebox{0.65}{\includegraphics{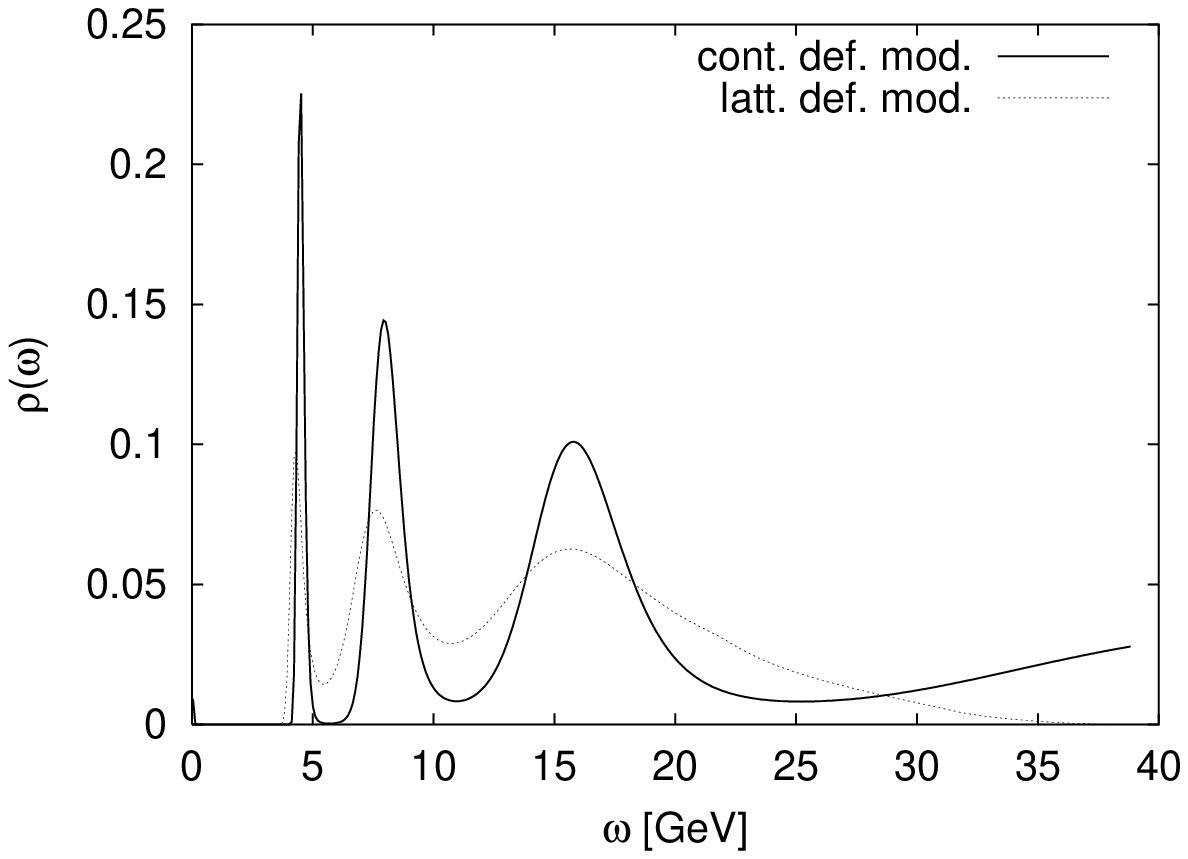}}
\scalebox{0.65}{\includegraphics{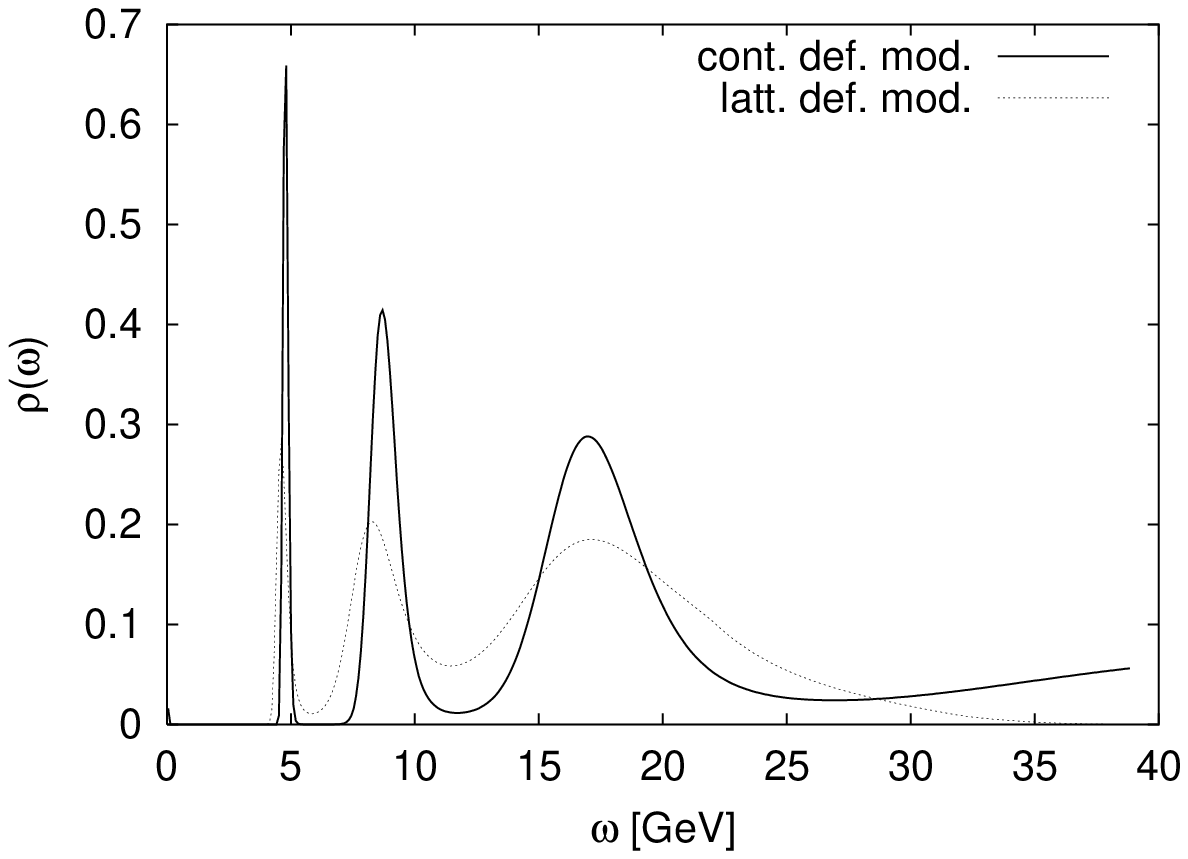}}
\caption{\label{fig.as719} 
Spectral functions for scalar (top) and axial vector (bottom) 
channels
reconstructed using the continuum and the lattice
free spectral function as a default model.
}
\end{figure}

Now let us discuss the scalar and axial vector channels.
In Fig. \ref{fig.as719} we show the spectral function in the scalar and
axial vector channels at $0.9T_c$ and set III using the free massive lattice
spectral function as a default model. Also shown there are the spectral
functions for free continuum default model from section IV for comparison.
As one can see the spectral functions show a stronger default 
model dependence than the pseudoscalar and vector ones, though the 
main features of the spectral functions are independent of the 
default model. The stronger default model dependence is probably due
to the fact that these correlators are noisier than the pseudoscalar
and vector ones. The default model dependence of the scalar and
axial vector spectral functions was found to be even
stronger for set II at $0.75T_c$.
If one uses the free massive spectral functions as default model at
$0.75T_c$  a large reduction in the peak height
of the $\chi_c$ state can be seen, though the position of the peak 
is roughly correct. This problem is also due to limited 
statistics. It turns out that the probability 
$P[\sigma|\alpha]=\exp(L-\alpha S)$, which is expected to
be strongly peaked around some $\alpha_{max}$ ideally (for
large statistics), has a long tail at large $\alpha$ values 
in this particular case, so that the peak structure seen
around $\alpha_{\rm max}$ is smoothened considerably when
one averages over $\alpha$.

\begin{figure}[htb]
\scalebox{0.65}{\includegraphics{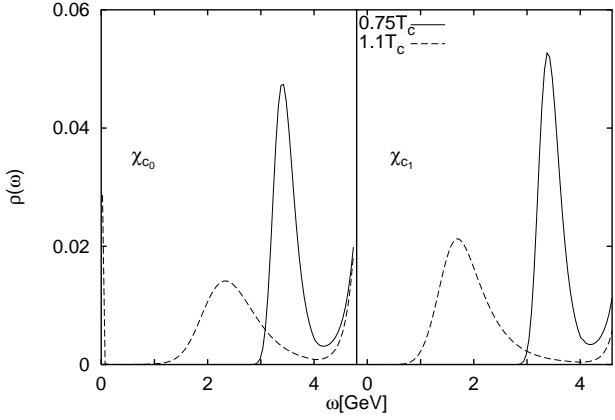}}
\caption{\label{fig.664nd8} Spectral functions from set II for 
scalar and axial vector with ${\rm N}_{\rm data}=8$ data points
at different temperatures.}
\end{figure}

Finally, we also want to address the issue whether the
serious system modification, possibly dissolution, of the
$\chi_c$ states at 1.12 $\tc$ seen in
Fig. \ref{fig.scax664} is an artefact of the smaller number
of data points at this temperature. In
Fig. \ref{fig.664nd8} we show the spectral functions in scalar and
axial vector channels at $0.75T_c$ and $1.12T_c$ using $N_{data}=8$
data points and the same temporal extent. The figure
clearly shows that the large change in the spectral
function between $0.75T_c$ and $1.12T_c$ is a physical
effect and not due to the small number of data points at
the higher temperature.

\begin{figure}[htb]
\scalebox{0.65}{\includegraphics{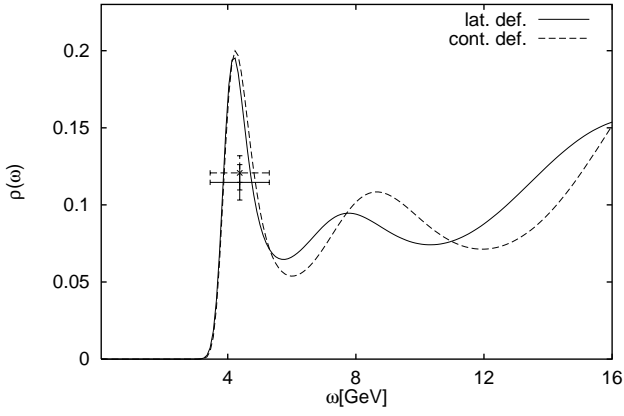}}
\caption{\label{fig.vc1.5free} 
The vector spectral function at $1.5T_c$ and set III using
the free continuum and lattice spectral function as a
default model.}
\end{figure}

So far all data above $T_c$ have been analyzed using
the high energy part of the corresponding spectral functions
below $T_c$. For set III at $1.5T_c$ we still have ${\rm N}_{\rm data}=12$
data points and not too small temporal extent. Therefore
here we can also reliably reconstruct the spectral function 
without providing precise information for the high $\om$
region. Figure \ref{fig.vc1.5free} show the spectral
function for $\jpsi$, reconstructed using both the free massless
continuum and the free massive lattice spectral functions
as the default model. 
Although some dependence on the default model is clearly seen,
the strong ground state peak is pretty stable and provides
further evidence that the $\jpsi$ survives well into the
deconfined gluon plasma.
 
\section{SPATIAL CORRELATORS AND SCREENING MASSES}
\label{sec.spatial}

Most of the early studies of the finite temperature meson spectrum 
concentrated on studies of correlators in one
of the spatial direction (for convenience, we will use the z-direction
in what follows). The corresponding masses are usually
referred to as the screening masses \cite{detar}. The
spatial correlators are related to the same spectral
function, but the relationship is a bit more complicated:
the momentum-projected correlators in the z direction are
given by \cite{review}
\beq
G(i \om_n, \vec{p}_T, z) = \int^\infty_{- \infty} {d p_z
\over 2 \pi} e^{i p_z z} \int^\infty_0 dp_0 {2 p_0
\sigma(p_0, \vec{p}_T, p_z) \over p_0^2 + \om_n^2}
\label{eq.spatcor} \eeq
where $\vec{p}_T$ is the transverse momentum in the
xy-plane and $\om_n$ is the Matsubara frequency. The
screening mass is obtained from the correlators projected
to zero transverse momentum and zero Matsubara frequency. 

While the screening masses, in general, are not directly
related to the spectrum of the finite temperature system,
it is possible to extract important informations from
them, in particular for examining particular model spectral
functions. In the presence of a stable bound
state, that gives a contribution 
$\sim \delta(p_0^2 - \vec{p}^2 - m^2)$ to the
spectral function, it can be easily seen from
Eq. (\ref{eq.spatcor}) that the screening mass is identical
to the pole mass of the state.

\begin{figure}[htb]
\scalebox{0.65}{\includegraphics{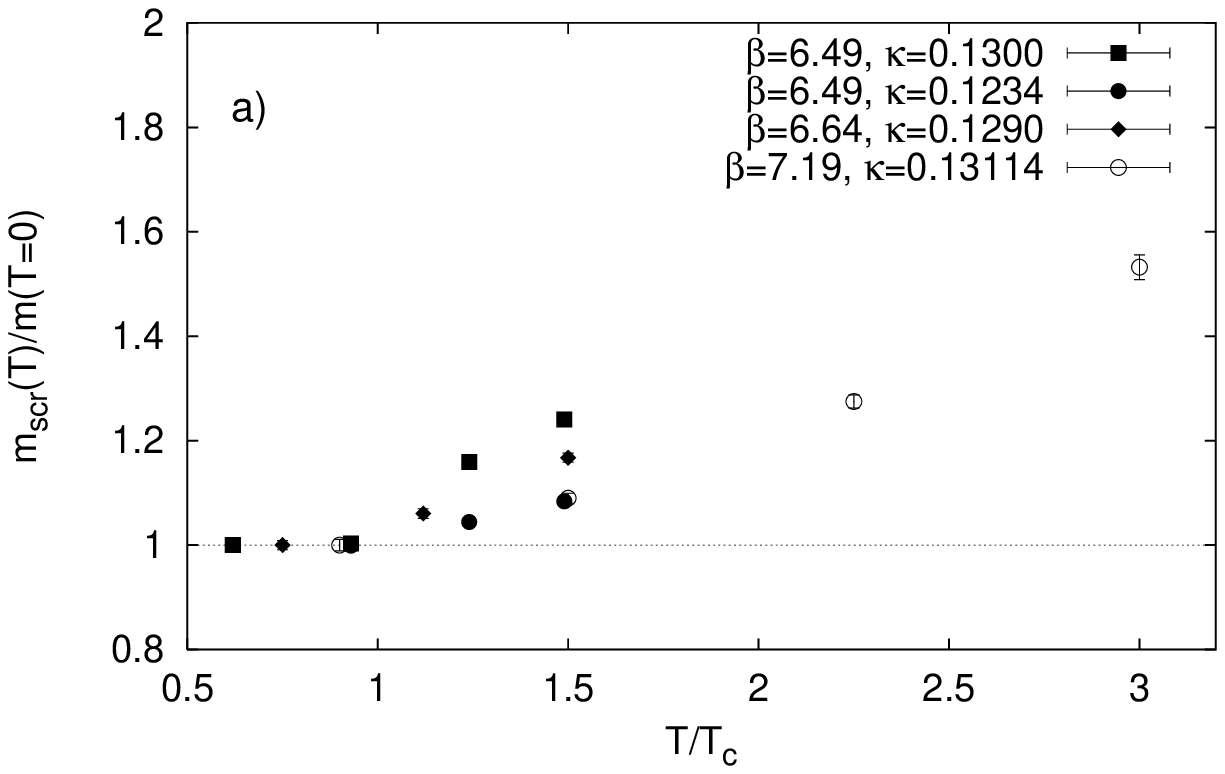}}
\scalebox{0.65}{\includegraphics{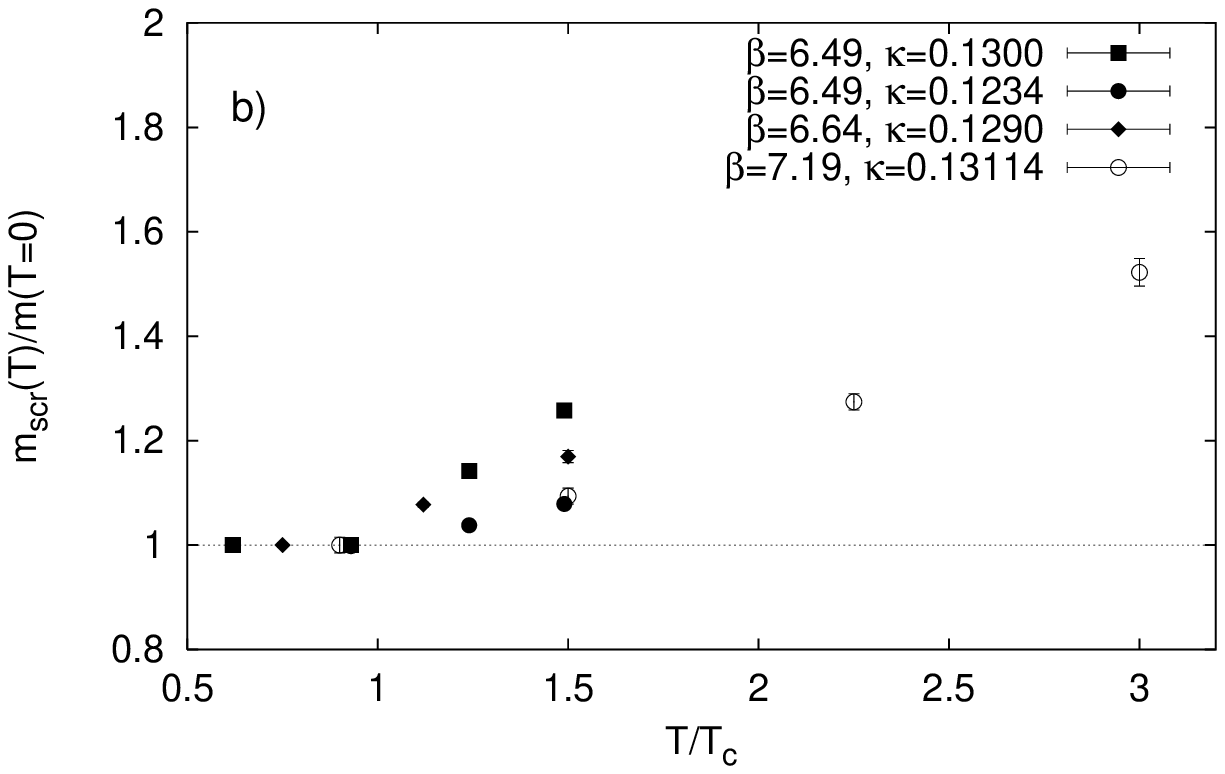}}
\caption{\label{fig.scrmass} Screening masses for the
a) vector and b) pseudoscalar channels for the different sets, 
normalized by the zero-temperature mass (see text).}
\end{figure}

Numerically, the study of screening masses is much simpler 
because of the large extent available in the z-direction, and 
one can obtain the screening masses from standard one-exponential
fits to the long distance part of the correlator. 
Fig. \ref{fig.scrmass} summarizes the temperature dependence of the 
vector and pseudoscalar screening masses for the different sets studied by us.
We have calculated
them from one-exponential and two-exponential fits, and in all cases
the two agree. In order to show
the relative changes in screening mass with temperature
we have normalized the masses with the zero-temperature
masses from Table \ref{tbl.parameters}, which, as mentioned
in Sec. \ref{sec.lattices}, were obtained from the
screening masses from the lowest temperature available to
us for each set. So the part below $T_c$ in
Fig. \ref{fig.scrmass} is just the normalization except in
Set I. From the figure, one makes these observations: 

1) The screening mass does not change upto
$T_c$ (or at least, till 0.93 $T_c$) for either channel. This observation is
in agreement with what is seen for light quarks and
glueballs. We also see that the screening masses below
$\tc$ are in agreement with the pole mass extracted from the
temporal masses, wherever such an extraction is possible
reliably.  

2) The screening masses show a definite temperature effect
above $\tc$, increasing steadily with temperature for both the 
channels. This effect is in sharp contrast to the temporal correlators,
which indicated little temperature effect upto
$\approx 1.5 \tc$, in particular for the pseudoscalar channel. 
Since for the pseudoscalar channel, we do not see any shift in the ground 
state peak position certainly upto 1.5 $\tc$, and possibly upto 2.25 $\tc$,
Fig. \ref{fig.scrmass} b) indicates that the pole and screening masses differ
from each other, even though a bound state is present in the spectral
function. For the vector case also, even though a small shift in the pole 
position may be present at 1.5 $\tc$, the shift of screening mass in 
Fig. \ref{fig.scrmass} a) is much larger and indicate a similar difference 
between the pole and screening masses. One possibility that one has to 
consider is that while there is a bound state in the spectral function above
$\tc$, it is not a $\delta$ function state. 
We investigated the possibility of
the screening mass change being caused by the state gaining
a width above $\tc$, by using a Breit-Wigner form as found in 
Ref. \cite{umeda02}, but our
conclusion is that the change in screening mass is too
large to be explained this way.

Another possible way that a screening mass can change
differently from the pole mass, even in the presence of a
bound state, is if the dispersion relation changes in the
thermal medium \cite{hashimoto}. Since in a thermal medium, the Lorentz
symmetry is lost, in general one would expect an
energy-momentum asymmetric self energy term, leading to a
dispersion relation 
\beq
\om^2(\vec{p},T) = m(T)^2 + \vec{p}^2 + \Pi(\vec{p},T).
\eeq
This will, in general, alter the screening mass from
the pole mass. For small momenta, one can make the simplifying 
assumption that 
the modified dispersion relation can be approximated by a
temperature-dependent mass and a temperature-dependent
``speed of light'' \cite{hashimoto}
\beq
\om^2(\vec{p},T) = m^2(T) + A^2(T) \vec{p}^2 
\label{eq.moddisp}
\eeq
For a delta-function state in the
dispersion relation, Eq. (\ref{eq.moddisp}) gives 
\beq
m_{\rm scr}(T) = m_{\rm pole}(T)/A(T).
\label{eq.scpole} \eeq
Such a modified dispersion relation can explain the discrepancy between
the pole and screening mass if, for example, $A(T) < 1$ at 1.5 $T_c$. 

\begin{figure}[htb]
\scalebox{0.65}{\includegraphics{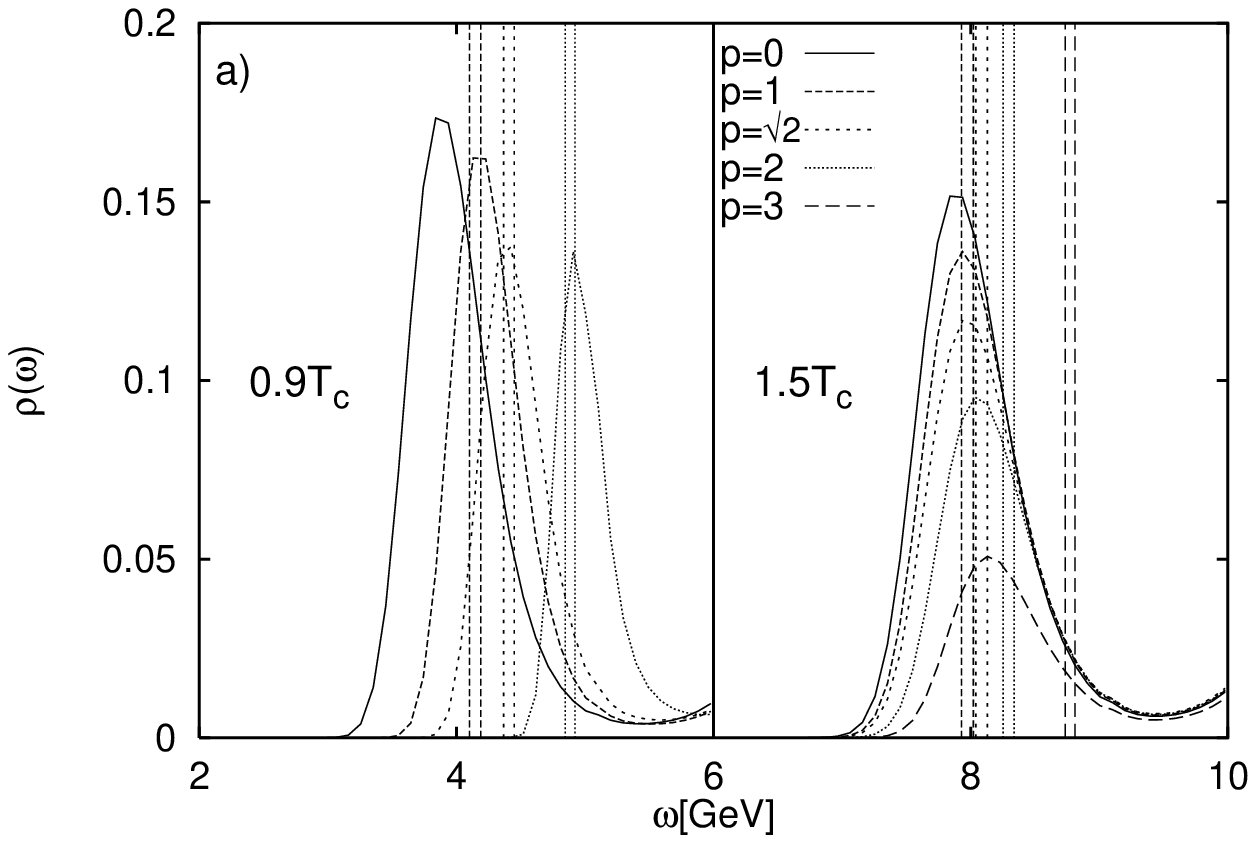}}
\scalebox{0.65}{\includegraphics{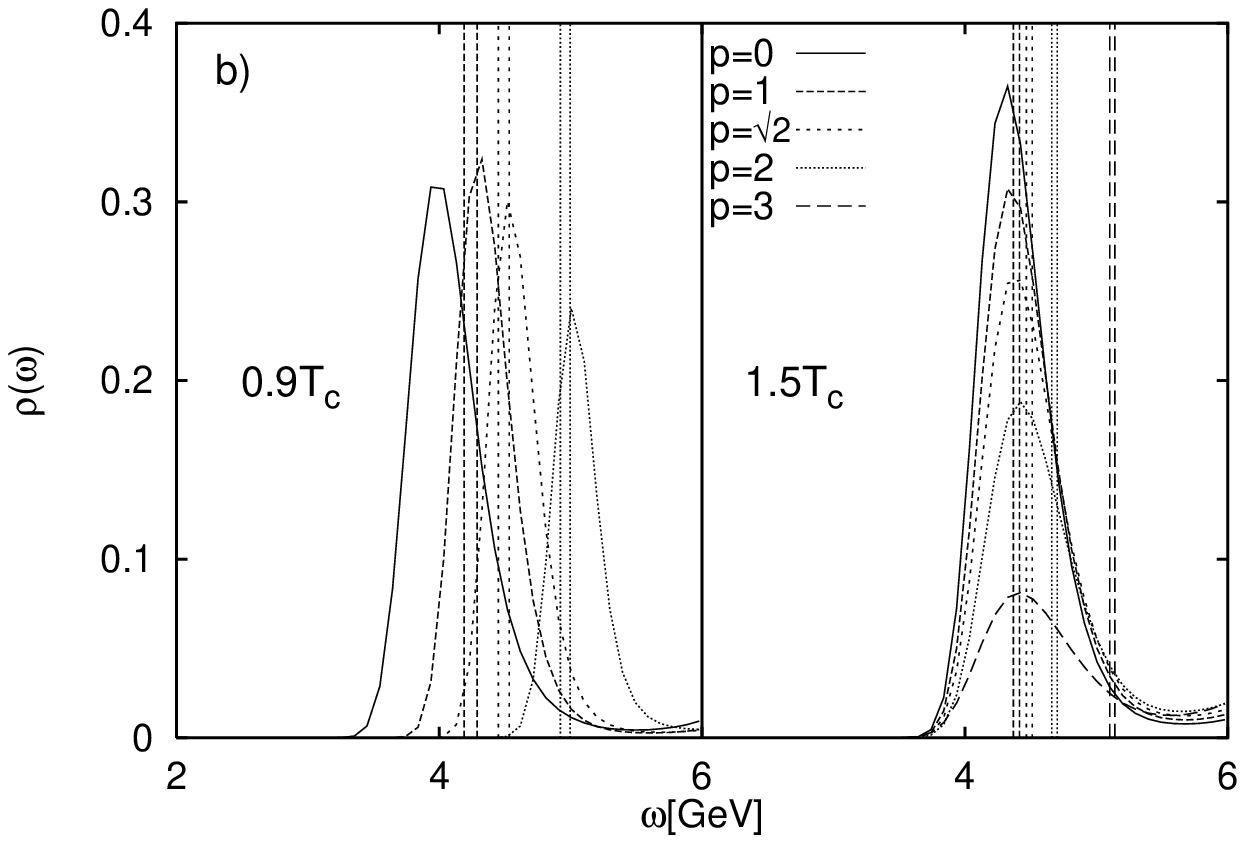}}
\caption{\label{fig.ftmom} Spectral functions for the a)
pseudoscalar and b) vector channel for a few non-zero
momenta, below and above $T_c$. The momenta are given in units of 
$2 \pi \over L a$, where $L a$ is the transverse dimension of 
the lattice.The vertical bands show the
expected peak positions for the non-zero momenta, assuming a
relativistic dispersion relation.}
\end{figure}

This scenario can be checked by constructing the spectral function from the 
temporal correlators for non-zero momenta. Eq. (\ref{eq.moddisp}) with 
$A(T) < 1$ will lead to a less-than-relativistic shift in the
spectral function peak with spatial momenta. Figure
\ref{fig.ftmom} a) shows the ground state peak for the
vector channel below and above $T_c$ for set III, and
Fig. \ref{fig.ftmom} b) shows the same for the pseudoscalar
channel. The vertical bands 
in both the figures show the expected peak positions for the non-zero momenta, 
assuming a relativistic shift from the peak at zero
momenta. This was obtained by using the lattice
form of Eq. (\ref{eq.moddisp}) with $A(T) = 1$,
\beq
\cosh \om(\vec{p})-1  = \sum_{i=1,2} (1- \cos p_i) +
{m^2 \over 2},
\label{eq.latdisp} \eeq
where all the variables $\om, p_i$ and $m$ are in units of inverse lattice
spacing. The width of the band reflects the uncertainty in
the zero momentum peak position due to the finite bin
size. The shift of the peak position at 0.9 $T_c$ is
seen to be consistent with the expected relativistic shift
in both the channels, in agreement with previous results that the Lorentz
symmetry is approximately valid upto close to $\tc$ \cite{npb,prd}. 
At 1.5 $T_c$, one clearly sees
a much smaller shift in the peak position, confirming the
medium modification of the dispersion relation discussed
above. A more quantitative analysis, to check whether the
shift in the screening mass is completely due to such a
modification of the dispersion relation, cannot, however,
be conclusively made due to the rather large uncertainty in
determining the peak position from a maximum entropy analysis.

\section{SUMMARY AND DISCUSSION}
\label{sec.summary}

In this work, we have conducted a detailed lattice study of the 
properties of the 1S ($\ec$ and $\jpsi$) and 1P ($\as$ and $\ax$) 
charmonia in hot gluonic medium, by studying the thermal correlators
in Euclidean time and extracting spectral functions from them.
We show that one can extract useful information about medium 
modifications of charmonia above $\tc$ by comparing the
correlators with those reconstructed from the spectral
function below $\tc$. We also show that by providing 
correct information about the structure of the high 
energy part of the spectral function, 
one can reliably extract the low energy part of the
spectral function at high temperatures from correlators
measured at a limited number of points.

In the hadronic phase, we find that the properties of all
the states remain unchanged at least upto temperatures of
0.93 $\tc$ as one approaches the deconfinement transition
temperature from below. This is similar to quenched studies
of light mesons and glueballs, and may not be true for full QCD.

As the temperature of the medium crosses $\tc$, significant
modifications of the $\chi$ states are observed. The
temporal correlators show large enhancement, and already at
1.1 $\tc$, the ground state peaks are not observed in the
spectral function, indicating that these states may
have dissociated already at this temperature. The
1S states behave quite differently. Little change is seen
in the correlators as one crosses $\tc$, and the spectral 
function shows a significant ground state peak till quite
high temperatures. No significant reduction
of either the peak strength or the mass of the state is
seen at least upto 1.5 $\tc$. At higher temperatures, the
peak weakens, though at 2.25 $\tc$ one still finds a
significant peak. (The apparent contradiction of this result with 
Ref. \cite{asakawa03} has been discussed in Sec. \ref{sec.abovetc}.)
Finally, at 3 $\tc$ we do not observe any significant peak 
any more, indicating that the state has
broadened and weakened so much that it is not meaningful to
treat it as a resonance any more.

These results have direct phenomenological implications, as 
$\rhw$ for the vector current is connected to the dilepton 
rate. Let us consider a 
hypothetical heavy ion collision experiment which forms equilibriated
gluon plasma, and whose temperature can be increased
gradually. As one crosses $\tc$, one expects to see a
reduction in the $\jpsi$ peak in the dilepton channel
caused by the disappearance of the excited states. (We would like to 
remind here that about $40\%$ of all $\jpsi$ produced in hadron
collisions are indirect and come from $\chi_c$ and $\psi^\prime$ states.)
The remaining $\jpsi$ peak then stays stable as one increases
the temperature further, showing little further reduction
at least upto 1.5 $\tc$. On further increasing the
temperature, the peak starts to ``melt'' gradually  by
weakening and possibly broadening. At 2.25 $\tc$ we see
$\sim 25 \% $ further reduction in the integrated strength
of the $\jpsi$ peak (for our quark mass). This process continues as one
increases the temperaure further, and by 3 $\tc$ the
$\jpsi$ peak is indistinguishable from the background.

It is reasonable to expect that the qualitative picture
remains the same as one introduces dynamical quarks. Below
$\tc$, there may be changes in the meson properties due to
the activation of the thermal pions, and it is possible
that the $\chi$ states dissociate even before $\tc$. For
$\jpsi$ itself, such thermal pions will have little effect,
and above $\tc$, the gradual melting picture described
above is likely to be still true qualitatively, with the 
thermal quarks causing only small quantitative changes.

These results are quite different from the earlier potential model
studies, which predicted the 1S charmonia to be dissolved
at $\sim 1.1 \tc$ \cite{karsch,digal}. Since the
appearance of our earlier studies \cite{lat02} some of the
potential model studies have been reanalyzed. It was
pointed out in Ref. \cite{shuryak} that Ref. \cite{karsch}
does not take into account a possible rise in the strong
coupling constant near $\tc$ which will lead to a bound
$\jpsi$ deep in the plasma. Also a very thorough and 
detailed study of the color singlet potential 
conducted in Ref. \cite{zantow}
shows that the color singlet free energy and potential
above $\tc$ (upto a few times $\tc$) behave very
differently from what was anticipated in Ref. \cite{digal} 
on perturbative grounds\cite{nadkarni}, and will probably
give rise to a considerably higher dissolution temperature for the
$\jpsi$. Be that as it may, it is not clear that a
potential model study can capture all the physics going
into the medium modification of $\jpsi$ like the collision
broadening due to thermal gluons. It will certainly be
interesting to have an analysis based on potential models
or other similar studies that can reproduce all the
features of the direct studies by us (and others), which also 
rules out any substantial reduction of the mass of the $\jpsi$ and
$\ec$ above $\tc$ from its zero temperature mass.

We will like to thank Edwin Laermann and Sven Stickan for
discussions. This research is funded by GSI under contract 
BI-KAR, and by DOE under contract DE-AC02-98CH10886.

\end{document}